\documentclass[twocolumn]{aastex63}

\usepackage{xcolor}
\usepackage{amsmath}
\usepackage{xcolor}

\submitjournal{}

\shorttitle{FRB 121102 persistent radio source}
\shortauthors{Chen et al.} 

\graphicspath{{./}{}}

\begin{document}

\title{A comprehensive observational study of the FRB 121102 persistent radio source}
\correspondingauthor{Ge Chen}
\email{gcchen@caltech.edu}

\author{Ge Chen} 
\affiliation{Cahill Center for Astronomy and Astrophysics, MC 249-17 California Institute of Technology, Pasadena CA 91125, USA}

\author{Vikram Ravi}
\affiliation{Cahill Center for Astronomy and Astrophysics, MC 249-17 California Institute of Technology, Pasadena CA 91125, USA}

\author{Gregg W. Hallinan} 
\affiliation{Cahill Center for Astronomy and Astrophysics, MC 249-17 California Institute of Technology, Pasadena CA 91125, USA}

\begin{abstract}

FRB 121102 is the first fast radio burst source to be spatially associated with a persistent radio source (QRS121102), the nature of which remains unknown. Motivated by the importance of the potential insights into the origins of FRBs, we present a detailed observational study of QRS121102 and its host galaxy. We constrain the physical size of QRS121102 by measuring its flux-density variability with the VLA in the Ku-band (12 to 18 GHz) and the K-band (18 to 26 GHz). Any such variability  would likely be due to Galactic refractive scintillation, and would require the source radius to be $\lesssim 10^{17}$ cm at the host-galaxy redshift.  We found the radio variability to be lower than the scintillation theory predictions for such a small source, leaving open the possibility for non-AGN models for QRS121102.  In addition, we roughly estimated the mass of any potential supermassive black hole (BH) associated with QRS121102 from the width of the H$\alpha$ emission line using a medium-resolution optical spectrum from the Keck Observatory.  The line width gives a velocity dispersion of $\lesssim 30$ km/s, indicating a supermassive BH mass of $\lesssim 10^{4 \sim 5}~M_{\odot}$.  We find the supermassive BH mass too low for the observed radio luminosity, and X-ray luminosity constraints, if QRS121102 were an AGN.  Finally, some dwarf galaxies that host supermassive black holes may be the stripped cores of massive galaxies during the tidal interactions with companion systems.  We find no nearby galaxy at the same redshift as the QRS121102 host from low-resolution Keck spectra, or from the PanSTARRS catalog.  In conclusion, we find no evidence supporting the hypothesis that the persistent radio source associated with FRB 121102 is an AGN. We instead argue that the inferred size, and the flat radio spectrum, favors a plerion interpretation. We urge continued broadband radio monitoring of QRS121102 to search for long-term evolution, and the detailed evaluation of potential analogs that may provide greater insight into the nature of this remarkable, mysterious class of object.  

\end{abstract}


\section{Introduction} \label{sec:intro}

Fast radio bursts (FRBs) are a class of transient phenomena wherein energies $\gtrsim10^{35}$\,erg are released on timescales $\ll1$\,s at radio wavelengths (e.g. \citealt{2007Sci...318..777L}).  The progenitor and emission processes of FRBs remain uncertain.  Hundreds of FRB sources have been reported, and over 20 of them have been found to repeat (e.g. \citealt{2021arXiv210604352T}).  Repeaters and non-repeaters are reported to show statistically different characters (luminosity, pulse width, temporal-spectral structures, etc.), yet it remains unclear whether or not they originate from two \textcolor{black}{distinct} populations.  FRB 121102 \textcolor{black}{was} the first repeater detected \citep{2016Natur.531..202S, 2016ApJ...833..177S}, and so far one of the most extensively studied FRB sources.  \textcolor{black}{The bursts are found to have a $\sim 160$-day periodicity \citep{2021MNRAS.500..448C}}, and the source has been localized within a star forming region of a low-metallicity dwarf galaxy at a redshift of 0.19273 \citep{2017ApJ...843L...8B, 2017ApJ...834L...8M, 2017ApJ...834L...7T}, \textcolor{black}{giving a luminosity distance of 971 Mpc (using the recent Planck results implemented in \textit{astropy}: $H_0 = 67.4~ \rm km~ s^{-1}~ kpc^{-1}$, $\Omega_{m}=0.315$, $\Omega_{\Lambda} = 0.685$; \citealt{2020A&A...641A...6P}).}

FRB 121102 is one of only two FRBs reported to be spatially associated with persistent radio emission of unknown origin \citep{2017ApJ...834L...7T,2021arXiv211007418N}. In the case of FRB 121102, the centroid of the persistent emission is within 12 mas (40 pc; 95\% confidence level) from the FRB source.  The emission shows a flat spectrum from $\sim$ 400 MHz to $\sim$ 6 GHz (flux density $\approx$ 200 $\mu$Jy), and decreases at higher frequencies ($166\pm 9$, $103\pm7$ and $66\pm7$ $\mu$Jy at 10, 15 and 22 GHz, respectively) \citep{2017Natur.541...58C,2020arXiv201014334R}.  It remains unresolved by very long baseline interferometry (VLBI) at 5 GHz, indicating a radius below $\sim$ 0.2 mas, or $\sim 10^{18}$ cm (0.35 pc) at the host-galaxy redshift \citep{2017ApJ...834L...7T}.  A flux-density amplitude modulation of $\sim30\%$ has been reported at 3 GHz \citep{2017Natur.541...58C}, consistent with refractive scintillation by the Milky Way interstellar medium (ISM) (e.g., \citealt{1986MNRAS.220...19R,1998MNRAS.294..307W}).  No X-ray counterpart has been detected with XMM-Newton and Chandra \citep{2017Natur.541...58C}.  

If we remain agnostic regarding models for the origin of FRBs, a compact radio source like that associated with FRB 121102, with a luminosity of $\sim2\times10^{29}$\,erg\,s$^{-1}$\,Hz$^{-1}$, would most likely be ascribed to AGN activity. Several AGN-like radio sources of similar luminosities have been reported to be hosted by dwarf galaxies (e.g. \citealt{2020ApJ...888...36R, 2019MNRAS.488..685M}). Although \citet{2017ApJ...834L...7T} has found the host-galaxy optical spectrum to be consistent with intense star formation based on the BPT diagnostics \citep{1981PASP...93....5B}, 
it is rare but not unheard of \citep[at the $\sim0.1\%$ level;][]{sbh+19} for galaxies classified as star forming according to BPT diagnostics to host radio-loud AGN. This may be substantially more common among the dwarf galaxy population: the majority of the \citet{2020ApJ...888...36R} sample of dwarf galaxies with optical spectra hosting candidate radio AGN are classified as star forming on BPT diagrams. The radio source associated with FRB 121102 is too compact and too luminous to be associated with star-formation activity \citep{2017Natur.541...58C}. Other possible origins include a supernova afterglow powered by interaction with a dense circum-stellar medium \citep{2021Sci...373.1125D}, the afterglow of a long-duration gamma-ray burst (e.g., \citealt{2003Natur.426..154B}), and an extreme pulsar wind nebula (PWN; e.g., \citealt{2018ApJ...868L...4M}).  

In this paper, we investigate the nature of the persistent radio source associated with FRB 121102 (``QRS121102'' hereafter)  using new data from the Karl G. Jansky Very Large Array (VLA) and the Low Resolution Imaging Spectrometer (LRIS) at the Keck Observatory.  We adopt the host redshift of 0.19273 \citep{2017ApJ...834L...7T} for all relevant calculations hereafter.  In Section \ref{sec:obs}, we describe the observations.  In Section \ref{sec:data}, we first measure the flux-density modulation of QRS121102 in the K-band (18 to 26 GHz) and the Ku-band (12 to 18 GHz), where refractive scintillation is expected to produce larger modulations than previously observed at 5 GHz.   We also separately investigate the hypothesis that QRS121102 is powered by a supermassive or intermediate-mass black hole (BH) using a medium-resolution LRIS spectrum, and evaluate whether or not the host galaxy belongs to a galaxy group using the low-resolution LRIS spectra.  In Section \ref{sec:discussion}, we first constrain the size of QRS121102 by comparing its flux-density modulation with that predicted by scattering theory.  In addition, we compare the AGN population with our dynamical BH mass estimation, stellar mass estimation, and radio and X-ray luminosity constraints.  We conclude in Section \ref{sec:conclusion} that an AGN hypothesis for QRS121102 is unlikely.

\section{Observations} 
\label{sec:obs}
  \subsection{VLA radio observations} \label{subsec: obs_VLA}

We have observed the persistent radio source (QRS121102) in the VLA K-band (18 to 26 GHz) and {Ku-band} (12 to 18 GHz) using the C configuration. The channel width was {2 MHz and the integration time was 3 s for the K-band observations and 2 s for the Ku-band observations.}  The observations include six epochs from 2017 May 29 to 2017 August 10 (Table \ref{ta:obs_VLA_epochs_fluxed}).  In each epoch, the observation started with a single scan of the primary calibrator 3C147 (Field 0) for flux scale and bandpass calibration, and then a few cycles of the phase calibrator 1 J0555+3948 (Field 1), phase calibrator 2 J0518+3306 (Field 2), and QRS121102 (Field 3) (Table \ref{ta:obs_VLA_targets}). 
  
\begin{deluxetable*}{c c c c c c c}
\tablecaption{VLA Observations and Results 
\label{ta:obs_VLA_epochs_fluxed}}

 \tablehead{
  \colhead{Obs.} & \colhead{Epoch} & \colhead{Band} & \colhead{{3C147 (f0)}} & \colhead{J0555+3948 (f1)} & \colhead{J0518+3306 (f2)} & \colhead{QRS121102 (f3)} \\
  \colhead{} & \colhead{(YYYY-MM-DD hh:mm)} & \colhead{} & \colhead{(Jy)} & \colhead{(mJy)} & \colhead{(mJy)} & \colhead{($\mu$Jy)} 
  }
  
 \startdata
  1 & 2017-05-29 16:09 & K & $1.994 \pm 0.010$ & $1956.5 \pm 2.5$ & $153.1 \pm 5.3$ & $56.6 \pm 8.2$ \\ 
  2 & 2017-06-03 16:43 & K & $2.010 \pm 0.013$ & $2135.8 \pm 3.4$& $143.9 \pm 2.5$ & $52.4 \pm 7.0$ \\
  3 & 2017-06-08 08:19 & U & $2.82 \pm 0.11$ & $2242 \pm 15$ & $200.2 \pm 1.7$ & $67 \pm 13$ \\
  4 & 2017-06-10 21:25 & U & \textcolor{black}{$2.857 \pm 0.010$} & $2565 \pm 30$ & $214.5 \pm 5.7$ & $118.0 \pm 5.5$ \\
  5 & 2017-06-11 16:32 & U & $2.799 \pm 0.018$ & $2431.2 \pm 6.4$ & $158.2 \pm 2.5$ & $82.1 \pm 6.4$ \\
  6 & 2017-08-10 12:31 & K & $2.050 \pm 0.026$ & $2148.4 \pm 7.5$& $193.6 \pm 5.4$ & $109 \pm 14$ \\
    & 2017-08-10 06:04 & U &  $2.804 \pm 0.026$ & $2217.7 \pm 6.0$ & $144.0 \pm 4.3$ & 63.5 $\pm 6.3$ \\ 
 \enddata 
\end{deluxetable*}

\begin{deluxetable*}{c c c c c c}
\tablecaption{VLA Targets
\label{ta:obs_VLA_targets}}

 \tablehead{
  \colhead{Field} & \colhead{Target} & \colhead{RA} & \colhead{DEC} & \colhead{Intention} & \colhead{Single Scan Duration} \\
  \colhead{} & \colhead{} & \colhead{} & \colhead{} & \colhead{} &  \colhead{(Minutes)} 
 }
 
 \startdata
  0 &  J0542+4951 (3C147) & $05^{\rm h}42^{\rm m}36.1^{\rm s}$ & $+49^{\rm o}51'07.2''$ & {Flux and Bandpass Calibrator} & 6 \\ 
  1 & J0555+3948 & $05^{\rm h}55^{\rm m}30.8^{\rm s}$ & $+39^{\rm o}48'49.2''$ & Phase Calibrator 1 \tablenotemark{a} & 1.5\\ 
  2 & J0518+3306 & $05^{\rm h}18^{\rm m}05.1^{\rm s}$ & $+33^{\rm o}06'13.4''$ & Phase Calibrator 2 \tablenotemark{b} & 1.5 \\ 
  3 & QRS121102 & $05^{\rm h}31^{\rm m}58.7^{\rm s}$ & $+33^{\rm o}08'52.5''$ & Science & 10 \\
 \enddata 
 \tablenotetext{a}{Used as the phase calibrator during our CASA imaging process.}
 \tablenotetext{b}{Treated as a science target during our CASA imaging process.}
\end{deluxetable*}

\begin{deluxetable*}{c c c c c c c}
\tablecaption{{Summary of Optical Observations}
\label{ta:obs_keck}}

 \tablehead{
  \colhead{Date} & \colhead{Instrument} & \colhead{Grating, Grism} & \colhead{$\lambda$} & \colhead{Slit} & \colhead{Resolution ($1 \sigma$)} & \colhead{Exposure}\\
  \colhead{(YYYY-MM-DD)} & \colhead{} & \colhead{(Red, Blue)} & \colhead{(Red, Blue; \AA)} & \colhead{} & \colhead{(\AA)} & \colhead{(Minutes)} 
 }
 
 \startdata
  2018-10-12 &  LRIS & 1200/7500, 400/3400 & NA & long 1.0'' & $\sim 1$ & 80 \tablenotemark{a} \\
  2017-01-26 &  LRIS & 400/8500, 600/4000 & 5462 $\sim$ 10318, 3122 $\sim$ 5603 & long 1.5'' & $\sim$ 4 & 50 \\ 
 \enddata 
 \textcolor{black}{\tablenotetext{a}{Each exposure is 20 minutes.  The total exposure time is 80 minutes when combining all four exposures, and 40 minutes after excluding the two exposures polluted by cosmic rays (see Section \ref{subsec: lris_high}).}}
\end{deluxetable*}

  \subsection{Keck optical observations}
  \label{subsec: obs_keck}

The optical spectra used in this work were obtained using Keck/LRIS. We obtained two types of observations. One, with medium spectral resolution, was used to measure the spectral width of the H$\alpha$ emission line associated with the FRB 121102 host galaxy. The other, with low spectral resolution, used two slit orientations to obtain spectra of galaxies immediately adjacent to the FRB 121102 host.

The medium resolution spectrum of the FRB 121102 host was obtained on October 12, 2018 using the 1.0'' longslit and the D560 dichroic.  The red side used the grating with 1200 lines/mm blazed at 7500 \AA~ and targeted the H$\alpha$ emission.

Low resolution spectra of the host and nearby sources were obtained on 2017 January 26 using the 1.5'' long slit and the 560 dichroic.  The grating used on the red side has 400 lines / mm blazed at 8500 \AA, \textcolor{black}{covering wavelength from about 5462 \AA~ to 10318 \AA, with a dispersion of 1.86 \AA/pixel.}  \textcolor{black}{We estimate the spectral resolution to be $\sim 4$ \AA~ (1$\sigma$) from the weighted average width of the four isolated skylines (5577.0~\AA, 5898.0~\AA, 6315.7~\AA, 7257.4~\AA)}.  On the blue side the grism has 600 lines / mm blazed at 4000 \AA, {covering wavelengths from 3122 ~\AA~ to 5603 ~\AA.} Standard stars were observed for the flux response calibration, arc-lamp spectra were obtained for the wavelength calibration, and bias frames were taken for the bias-subtraction.  The flat field was generated using  dome flats.

\section{Data Analysis and Results}
\label{sec:data}

  \subsection{VLA flux density measurements}
  \label{subsec:results_radio} 
  
In this section, we describe how we measure the radio flux density of \textcolor{black}{QRS121102} in each \textcolor{black}{epoch}. 
  
The visibility data \textcolor{black}{were calibrated and imaged using CASA following the standard procedures (task names shown in parentheses).\footnote{The Common Astronomy Software Applications (CASA) is a software package developed by the NRAO.}  In the calibration, the VLA antenna positions were updated (\textit{gencal}).  The primary calibrator 3C147 was used to find the absolute scale of the gain amplitudes by referring to the standard flux density (\textit{setjy}), and to correct for the instrumental delay and the complex antenna-response variation with frequency (\textit{bandpass}).  The complex gain solutions (both amplitude and phase) were obtained from the phase calibrator 1, J0555+3948, that is $\sim 8^{\rm o}$ apart from QRS121102 (CASA task \textit{gaincal}), and the gain amplitudes were properly-scaled using the absolute scale just obtained using 3C147 (\textit{fluxscale}).  All calibration solutions were then applied (\textit{applycal}) to QRS121102, as well as the fainter phase calibrator 2, J0518+3306 ($\sim 2^{\rm o}$ away from QRS121102)}.  After calibration, potential radio frequency interference (RFI) and internally generated signals were removed by flagging out narrow-banded spikes in the spectra (\textit{flagdata}).  \textcolor{black}{In epoch 3, one scan of QRS121102 was removed as it was several times brighter than the others (within 30 minutes apart) due to strong RFI, or perhaps a passing cloud.} 
  
  The calibrated visibility data were binned to $20$ s and $500$ kHz to speed up \textcolor{black}{synthesis} imaging using the CASA task \textcolor{black}{(\textit{split})}.  The data were gridded and Fourier transformed and the \textcolor{black}{synthesized} beam was deconvolved \textcolor{black}{(\textit{tclean})}.  {The FWHM of the synthesized beam was $\sim 1.5''$ (major axis) $\times 1.2''$ (minor axis) in the Ku-band and $\sim 1.3''$ (major axis) $\times 0.9''$ (minor axis) in the K-band} in C configuration.  Our images were created using a small cell size of $0.1''\times 0.1''$ and a Briggs robust weighting of 0.5 \citep{1995AAS...18711202B}.  
  
\begin{figure*}
\gridline{\fig{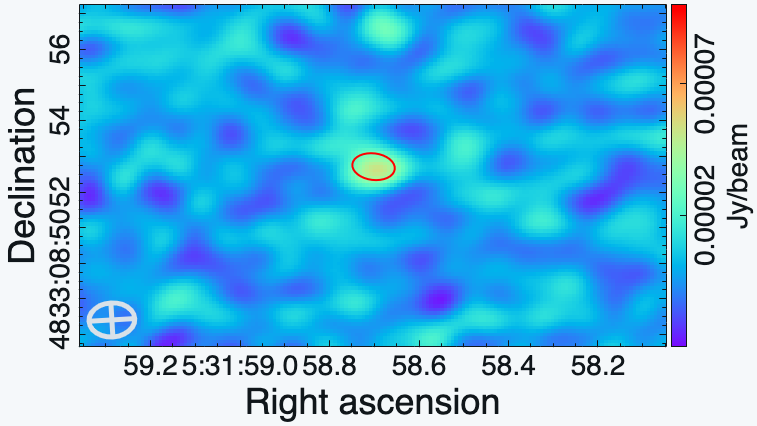}{0.33\textwidth}{Epoch 1 (K)}
          \fig{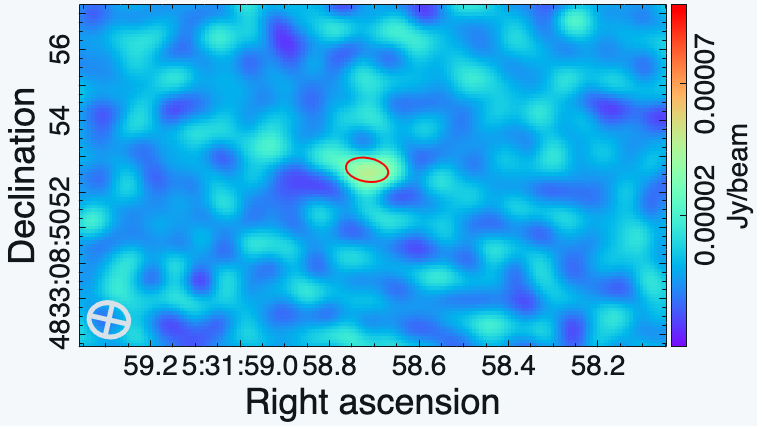}{0.33\textwidth}{Epoch 2 (K)}
          \fig{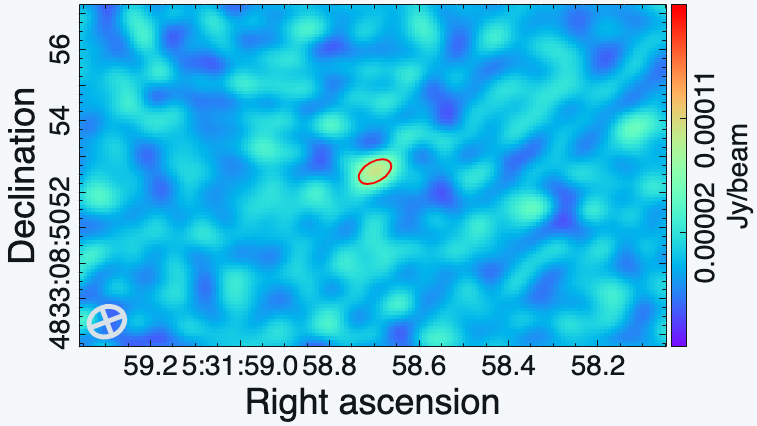}{0.33\textwidth}{Epoch 6 (K)}}
\gridline{\fig{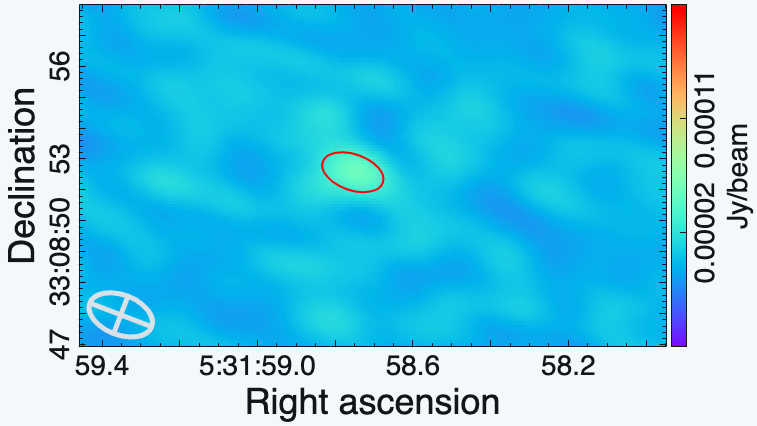}{0.33\textwidth}{Epoch 3 (U)}
          \fig{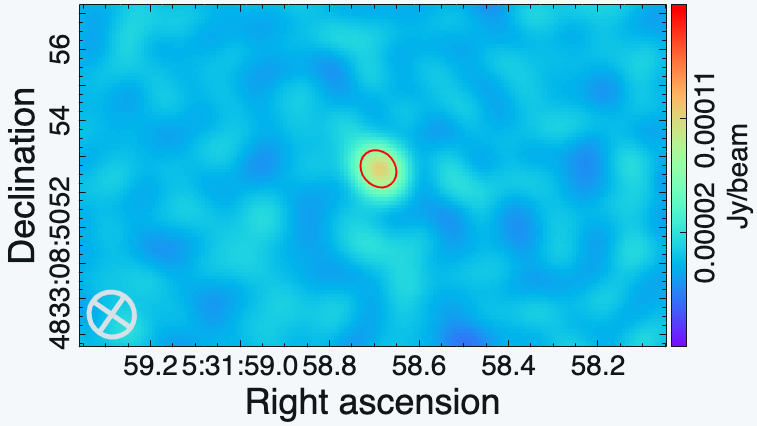}{0.33\textwidth}{Epoch 4 (U)}
          \fig{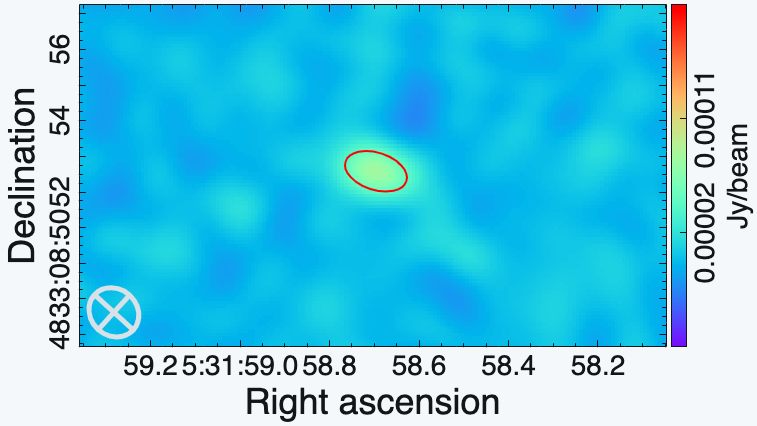}{0.33\textwidth}{Epoch 5 (U)}}
\gridline{\fig{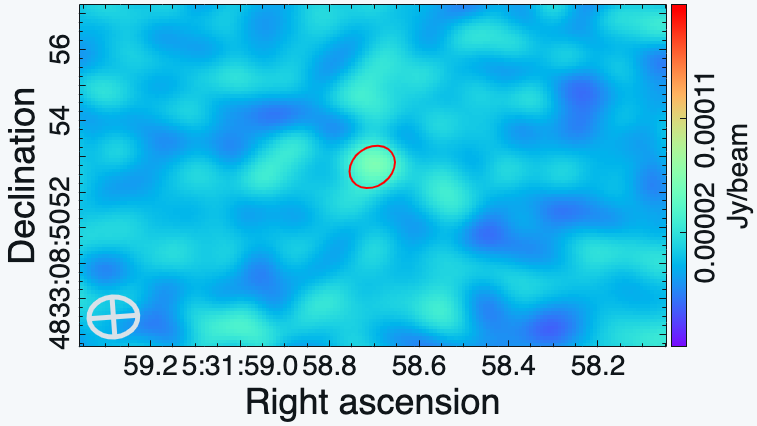}{0.33\textwidth}{Epoch 6 (U)}
          }
\caption{VLA images (in J2000 coordinates) of QRS121102 in seven epochs, with band indicated in parentheses.  The color scale represents flux density in Jy / beam (see color bar).  The open light gray circle (on the bottom left of each image) shows the synthesized beam size ($1 \sigma$) and the red circle shows the 1$\sigma$ 2D Gaussian fitting results convolved with the synthesized beam.  {The position angles of the best fit results are consistent with those of the clean beams within 1$\sigma$ in all cases where the source is unresolvable.} } 
\label{fig:VLA_images}
\end{figure*}

\begin{figure}
    \includegraphics[width=0.5\textwidth]{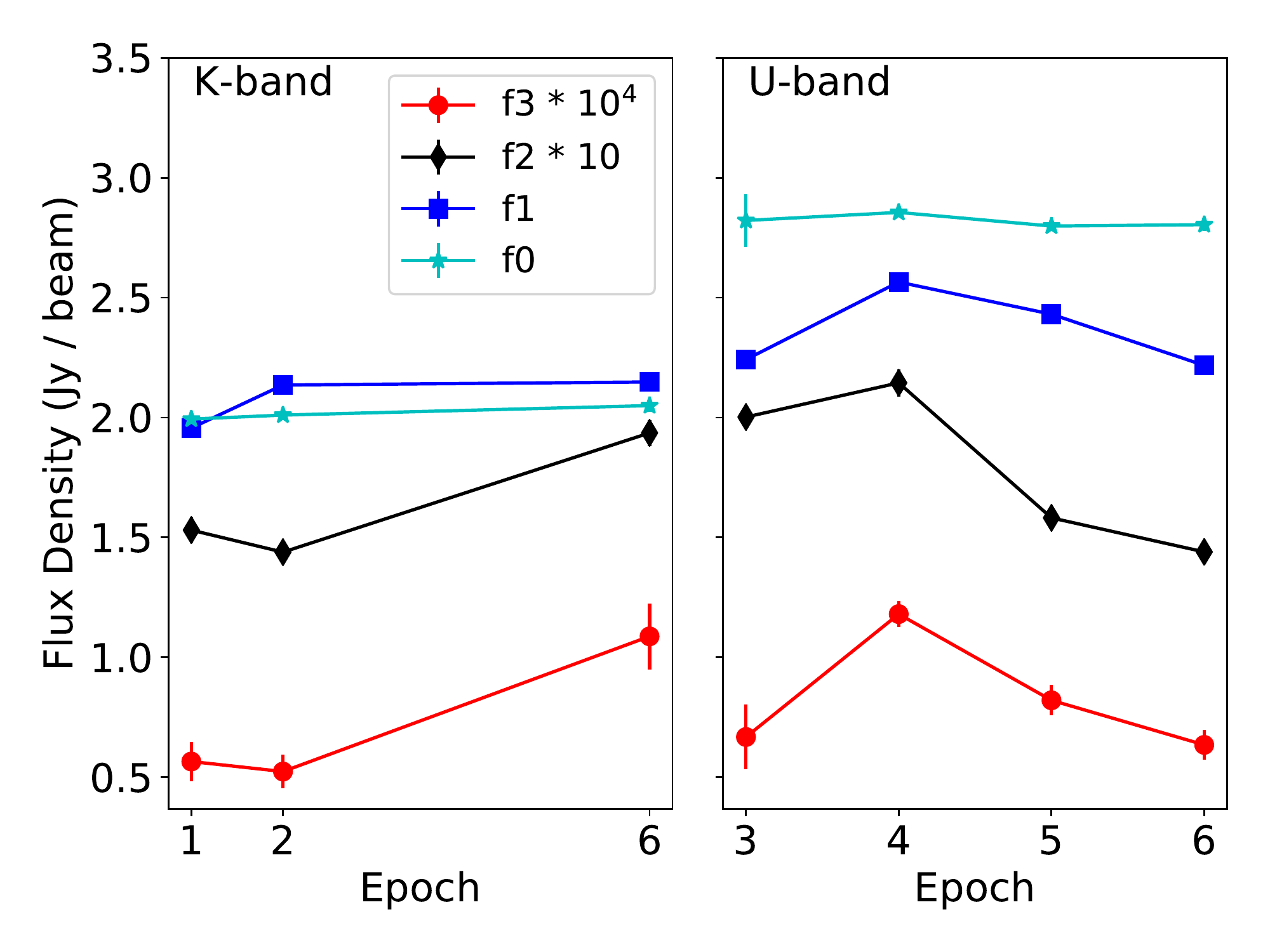}
    \caption{Flux density light curve of the K-band (left) and the \textcolor{black}{Ku-band} (right).  f3, f2, f1 and f0 represents the flux density of QRS121102 (solid red circle), the phase calibrator 2 (solid black diamond), the phase calibrator 1 (solid blue square) and the prime calibrator (cyan stars), respectively.  f3 and f2 are scaled by $10^{4}$ and 10 times for display.  {Details of the observations are shown in Tables \ref{ta:obs_VLA_epochs_fluxed}, \ref{ta:obs_VLA_targets}.} }
    \label{fig:lcv}
\end{figure}

\begin{figure}
    \includegraphics[width=0.5\textwidth]{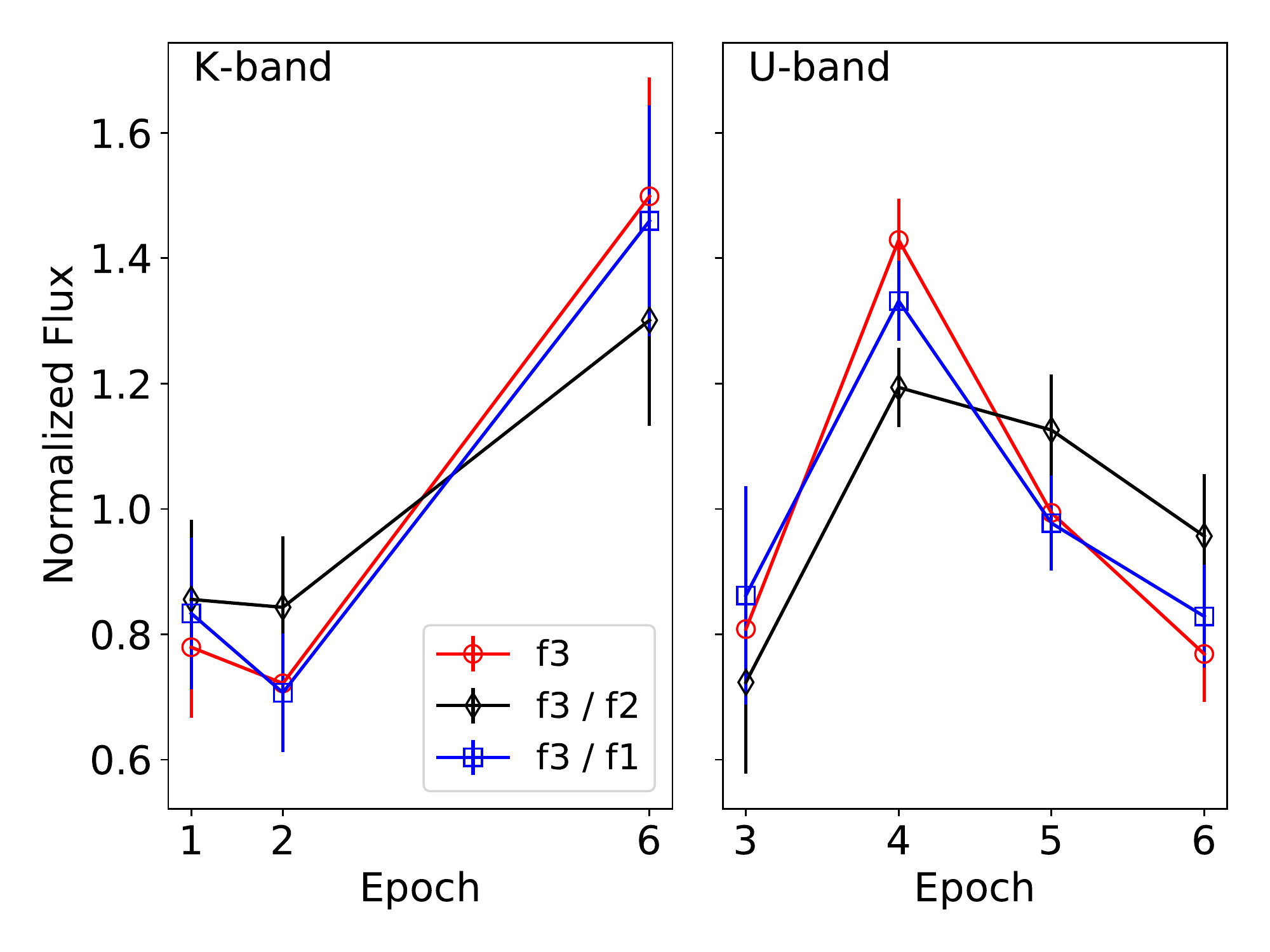}
    \caption{Normalized (and scaled) flux density light curve of the K-band (left) and the \textcolor{black}{Ku-band} (right).  f3, f2 and f1 represents the flux density of QRS121102, the phase calibrator 2 and the phase calibrator 1, respectively.  Red open circles (f3) show the normalized flux density of QRS121102, black open diamonds (f3/f2) show the flux density of QRS121102 divided by those of the phase calibrator 2 and then normalized to average at unity, and blue open squares (f3/f1) are the flux density of QRS121102 divided by those of the phase calibrator 1 and then normalized to average at unity.}
    \label{fig:lcv_norm_scaled}
\end{figure}

  We measured the \textcolor{black}{flux density} of each source from the images by fitting each with a 2-D Gaussian model (\textit{imfit}; Fig.\ref{fig:VLA_images}).  {We obtained the flux density of those point sources (based on the criterion implemented by the CASA \textit{imfit} task) from the peak, and of the others from the integrated flux density within each fitting region.  Uncertainties were calculated by
   propagation of errors of the 2D fitting model.}  Table \ref{ta:obs_VLA_epochs_fluxed} lists the results and Fig. \ref{fig:lcv} shows the flux density light curves in both bands.  
  
  The angular size of QRS121102 has been reported to be \textcolor{black}{under} $\sim$ 0.2 mas at 5 GHz as measured by the Very-long-baseline interferometry (VLBI) \citep{2017ApJ...834L...8M}.  It is expected to be unresolved in our observations, \textcolor{black}{where the} beam FWHM is $\sim$ 1.4'' in the \textcolor{black}{Ku-band} and $\sim$ 0.9'' in the K-band.  In our images, the 2D Gaussian fitting results show that QRS121102 is a point source in all but epoch 3, where the image is marginally resolved as $1.13 \pm 0.61$ arcsec along the major axis and $0.49 \pm 0.22$ arcsec along the minor axis (FWHM, deconvolvd from beam).  This is likely due to the remaining phase errors in the calibration solution, as the image of phase calibrator 2 J0518+3306 appears highly distorted.  The errors were removed by {phase-only self-calibration in field 2, but not in field 3, for reasons outlined below.}
  
  
 In field 3, our observations of QRS121102 were affected by a bright source, NVSS J053153+331014, that is $\approx$ 1.8' (twice of the primary beam FWHM in the Ku-band) away from the pointing center (QRS121102) and over one order-of-magnitude brighter {($\approx$ 3 mJy in flux density before correcting for primary-beam attenuation)}.  To reduce the associated errors, we performed self-calibration (phase only) for NVSS J053153+331014 \textcolor{black}{and QRS121102 simultaneously}.  We also tried to remove the \textcolor{black}{flux density} contribution from NVSS J053153+331014 modeled from self-calibration, and then {subtract the model visibility from the corrected visibility data.}  \textcolor{black}{Neither attempt made a significant improvement on the image of QRS121102, since self-calibration failed to correct the beam model error far away from the pointing center}.

  \textcolor{black}{In addition, there might be remaining calibration errors since the flux densities of the two phase calibrators, J0555+3948 and J0518+3306, also vary by $\sim 4\%$ and $\sim 10\%$ {throughout} the epochs in each band (Fig. \ref{fig:lcv}).  The variations are {likely not intrinsic to the source} for two reasons.  (1) J0555+3948 has been reported to vary by $2.0\%$ on a time scale of 251 days at 33 GHz and $3.4\%$ on a time scale of 293 days at 16 GHz \citep{2009MNRAS.400..995F}. It is unlikely to show a greater variability on a timescale of days, as been observed in this work (e.g. epochs 3, 4, 5).  (2) More importantly, the flux density light curve of QRS121102 shows a moderate positive correlation with that of J0555+3948, and a strong positive correlation with that of J0518+3306}, with a correlation coefficient of 0.67 and 0.91, respectively.  To reduce potential calibration errors, in each epoch we re-scaled the \textcolor{black}{flux density} measurements of QRS121102 by those of the two phase calibrators (Fig. \ref{fig:lcv_norm_scaled}).  We {adopt} the \textcolor{black}{flux density} scaled by J0518+3306 (phase calibrator 2) thereafter, {since it has a smaller angular separation from QRS121102}. 

\subsection{Keck/LRIS Medium Resolution Spectral Analysis} 
\label{subsec: lris_high}

\begin{figure}
\gridline{\fig{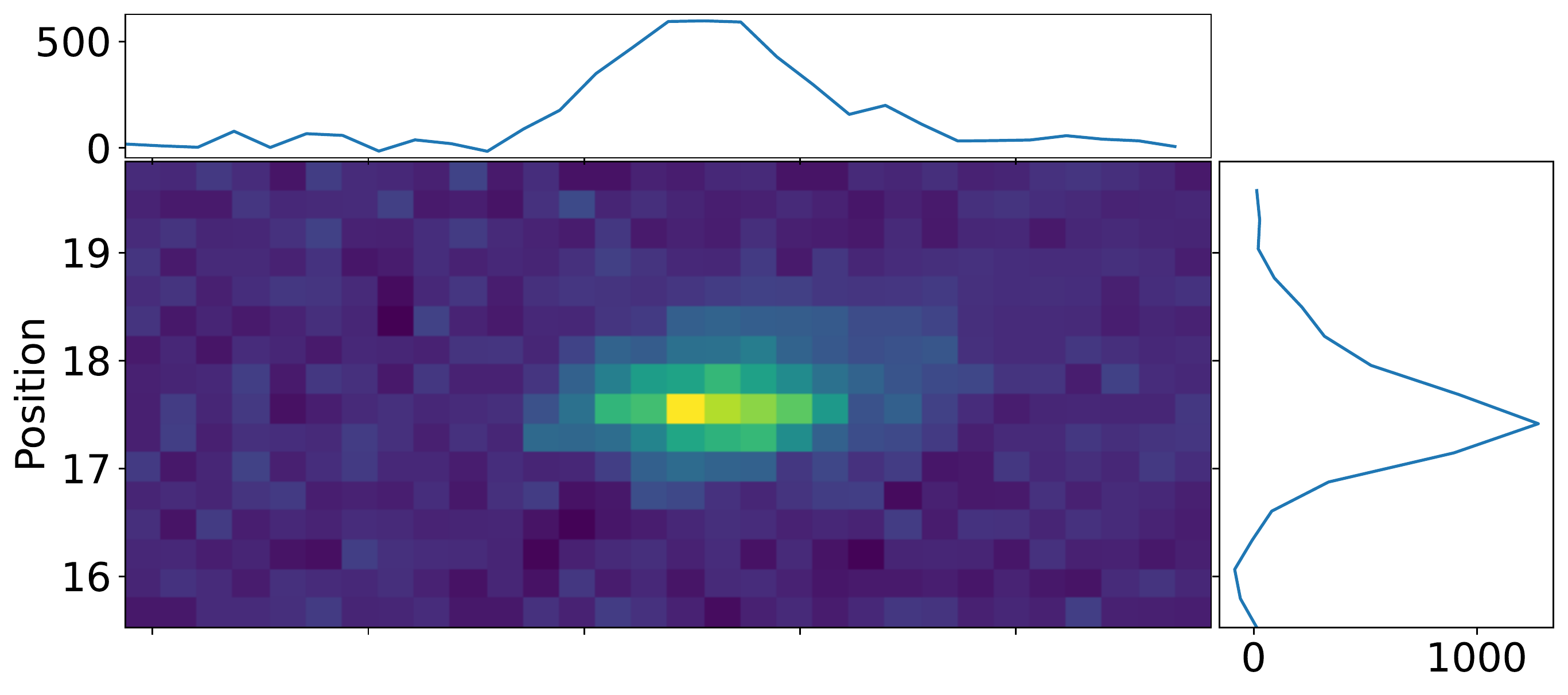}{0.5\textwidth}{}}
\gridline{\fig{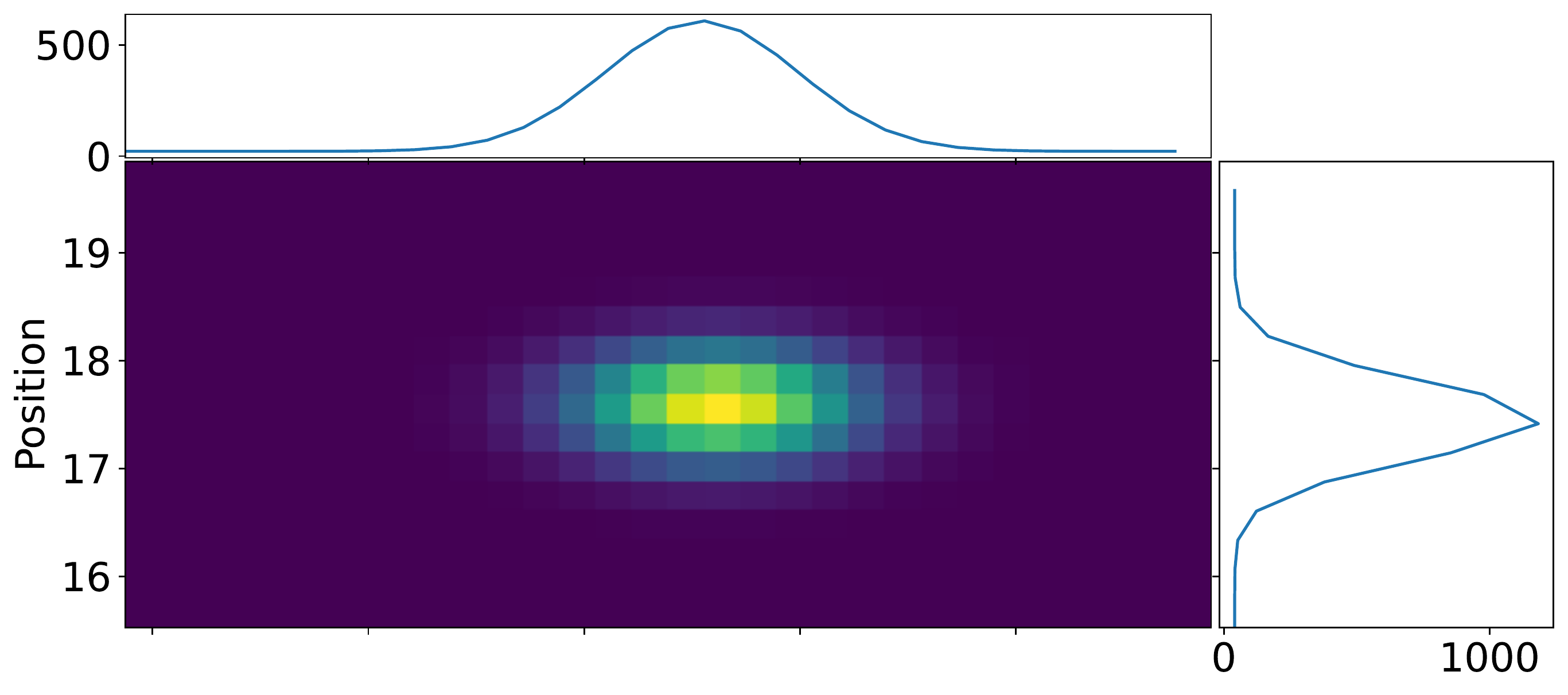}{0.5\textwidth}{}}
\gridline{\fig{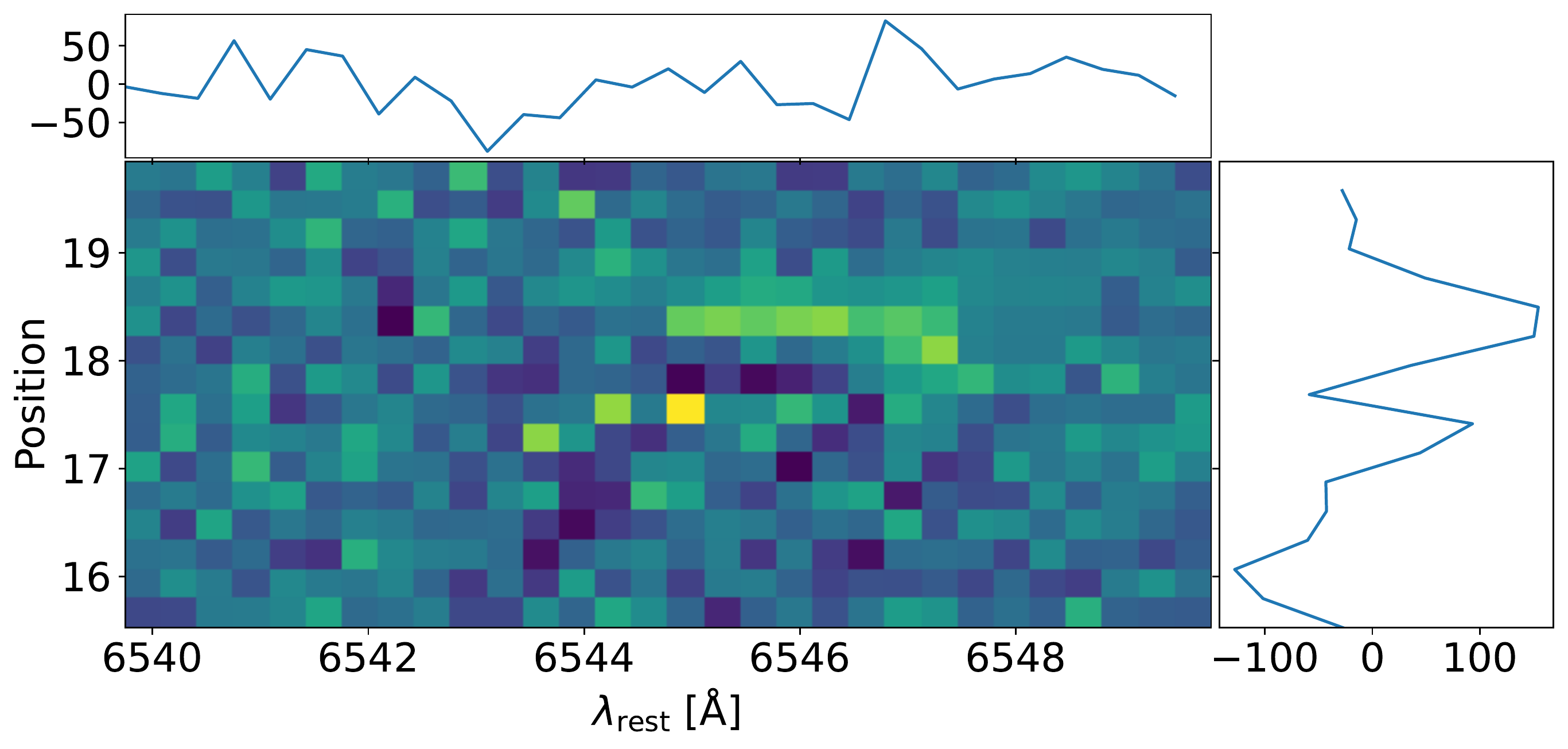}{0.5\textwidth}{}}
\caption{2D medium resolution LRIS spectrum of the host galaxy of FRB 121102.  Wavelength has been converted to the rest frame using the previously reported redshift of $z=0.19273$ \citep{2017ApJ...834L...7T}.  Top to bottom panels are the observed 2D spectrum, the 2D fitting model and the residual (data$-$model), respectively.  On the top and right of each panel are plots of the data collapsed along the wavelength axis and position axis, respectively.} 
\label{fig:2D_spec}
\end{figure}

\textcolor{black}{The LRIS data were processed using the LRIS automated reduction pipeline (LPipe; \citealt{2019PASP..131h4503P}) following standard procedures: subtract the bias, make flat fields and apply flat-field correction, remove cosmic ray pixels, model and remove sky lines, perform wavelength calibration by referring to the arc lamp spectra, and determine the flux response function by observing standard reference stars.  The processed 2D spectrum targets the H$\alpha$ emission.}

The width of the H$\alpha$ emission line is produced by multiple effects \citep{1979rpa..book.....R}-- 
  \begin{itemize}
    \item dynamical velocity dispersion due to gravity,
    \item instrumental broadening,
    \item random thermal motions,
    \item natural broadening,
    \item turbulent motions.
  \end{itemize} 
  In this section, \textcolor{black}{we test the hypothesis that QRS121102 is associated with an AGN by using the velocity dispersion to roughly constrain the mass of a potential supermassive BH.  The velocity dispersion is estimated from the H$\alpha$ line width.}

  The \textcolor{black}{H$\alpha$} emission line width of the host is determined by fitting the rest-frame 2D spectrum with a 2D Gaussian function whose rotation angle is fixed at zero, plus a constant offset. We convert the observed 2D spectrum of the host into the rest-frame wavelength using a previously reported redshift of 0.19273 \citep{2017ApJ...834L...7T}.    \textcolor{black}{The width (1$\sigma$) of the emission line is $0.9316 \pm 0.0026$ \r{A} when combining all four exposures, and $0.8947 \pm 0.0015$ \r{A} after removing the two exposures polluted by the nearby cosmic rays (Fig. \ref{fig:2D_spec}). A bright pixel at the center of the H$\alpha$ emission is seen in the residual (bottom panel of Fig. \ref{fig:2D_spec}), but no extended structure that might impact the emission width result is found.}

  We then determine the width of the instrumental broadening effects by collapsing the 2D spectrum into 1D and fitting each of the three isolated sky lines in the field with a 1D Gaussian \textcolor{black}{function} plus an offset.  The \textcolor{black}{instrumental} spectral broadening width is given by the weighted mean of the three sky lines'  1-$\sigma$ widths ($0.638725 \pm 0.000039$ \AA).  The natural broadening width of the \textcolor{black}{H$\alpha$} line is known to be 0.46 m\r{A} (e.g. \citealt{1979rpa..book.....R}).  The 1-$\sigma$ width of the line is \textcolor{black}{$0.6265 \pm 0.0021$ \r{A} after quadratically removing the instrumental and natural broadening effects.}
  
 The rest frame line width of a Maxwellian velocity distribution (i.e. Gaussian along the line-of-sight) is given by \citep{1979rpa..book.....R}:
  \begin{equation}
      \frac{\Delta \nu}{\nu_0} = \frac{1}{c} \left(\sigma_v^2 + \frac{2 k T}{m_{\rm H}} + v_{\rm turb}^2 \right)^{1/2}
  \end{equation}
  Here, $\Delta \nu$ Hz is the rest-frame line width (1$\sigma$) in frequency and  $\nu_0=4.57\times10^{14}$ Hz is the frequency of the \textcolor{black}{H$\alpha$ emission in vacuum}.  $\sigma_v$ is the velocity dispersion due to gravity, $\sqrt{2 k T/m_{\rm H}}\approx 12.8$ km/s is the most probable thermal velocity assuming a gas temperature of $10^4$ K (e.g. \citealt{2011piim.book.....D}).  $v_{\rm turb}$, the turbulent velocity, is weakly constrained to be $\lesssim 10^3$ km/s as inferred from the scattering measurements of FRB 121102 (Table 2 of \citealt{2021arXiv210711334S}).  We ignored its contribution and find an upper limit of $\sigma_v \lesssim 30 \rm~km~s^{-1}$.

Assuming that the FRB 121102 host galaxy has a central BH, we estimate its mass using the empirical M-$\sigma$ relation reported in recent literature. 
\textcolor{black}{The BH mass is $7.8^{+8.2}_{-5.2} \times 10^4 M_{\odot}$ using the relation derived from a sample of 88 AGN covering a stellar velocity dispersion $\sigma_*$ of 30 -- 268 km s$^{-1}$ (\citealt{2006ApJ...641L..21G}, error bars calculated from the intrinsic scatter found in the relation).  A consistent BH mass of $8.8^{16.2}_{-5.8} \times 10^4 M_{\odot}$ is found using the relation based on 93 low-mass active galaxies (\citealt{2011ApJ...739...28X}, $\sigma_*$ from 31 to 138 km/s).  Other reports show similar results, though an extrapolation of the M-$\sigma$ relation is required as the sampled objects cover higher $\sigma_*$: the mass is $4.2^{+7.5}_{-2.7} \times 10^4 M_{\odot}$ (error bars from the intrinsic scatter in the relation) based on a sample of 49 BH mass dynamical measurements in spiral galaxies, S0 galaxies and elliptical budges ($\sigma = $ 67--385 km s$^{-1}$; \citealt{2009ApJ...698..198G}),  and $\sim 10^4 M_{\odot}$ from 72 similar objects ($\sigma = $ 75 -- 347 km s$^{-1}$; \citealt{2005SSRv..116..523F, 2013ApJ...764..184M}).} Finally, we note that the measured velocity dispersion is lower than any of those measured from a sample of 35 tidal disruption events host galaxies ($ \sigma > 43$ km/s) reported by \citealt{2019MNRAS.487.4136W, 2017MNRAS.471.1694W}.  This suggests that the BH mass in the FRB 121102 host is lower than the BH masses of the tidal disruption event galaxy sample.

We consider two potential errors in our BH mass estimation. \textcolor{black}{First, the H$\alpha$ velocity dispersion measurement may not be suitable for the dynamical analysis.  The BH mass is usually estimated from the stellar velocity dispersion measured from multiple absorption lines from an optical or IR spectrum.  In our observation, these stellar absorption line widths were unavailable due to the limited SNR.  Instead, we infer the velocity dispersion from the line width of a single compact H$\alpha$ emission region (radius $< 0''.24$ at 1$\sigma$, \citealt{2017ApJ...844...95K}) that is offset from the stellar continuum centroid of the galaxy by $0''.29 \pm 0''.05$ \citep{2017ApJ...834L...7T}.  The H$\alpha$ line width reveals the dynamics of the partially ionized warm star forming gas formed in discrete clouds.  If the gas pervades in the galaxy, it is expected to have larger velocity dispersion than the stars due to turbulent motions and provide an upper limit to the BH mass.  However, the H$\alpha$ region is isolated to one part of the host and may not represent the global gas dynamics in the host.  This could lead to systematic errors as reported in the dynamical analyses of galaxies with irregular gas and dust distributions \citep{2002PASP..114..137H}.}

\textcolor{black}{However, the dynamical mass implied by the H$\alpha$ velocity dispersion is comparable to the stellar mass of the FRB 121102 host galaxy inferred using the optical/IR spectral energy distribution.  We assume that the  system is virialized for an order-of-magnitude estimation.}  In an ellipsoid, the kinetic energy is dominated by random motions.  The virial theorem gives a total stellar mass of $M_{*} \sim \sigma_v^2~R_{\rm eff}/G \sim 10^8 M_{\odot}$, adopting a half-light-radius of \textcolor{black}{$R_{\rm eff} = 0.68$ kpc \citep{2017ApJ...843L...8B}}.  In a pure rotational disk (e.g. disk of a spiral galaxy), a comparable value is expected.  The inferred mass is consistent with the stellar mass reported by \citet{2017ApJ...843L...8B} from a spectral energy distribution fit ($(1.3\pm0.4)\times10^{8}M_{\odot}$). This suggests that the velocity dispersion may be useful for the dynamical BH mass estimation. We also note that QRS121102 is spatially associated with the H$\alpha$ emission region.

\textcolor{black}{Second, the M-$\sigma$ relations could be less reliable at our velocity dispersion for two reasons.  (1) Most reports derive the empirical M-$\sigma$ relation based on a sample of more massive galaxies with larger stellar velocity dispersions.  It is unclear how well the relation extends to lower velocity dispersions.  (2) Even in the few reports that cover velocity dispersions down to 30 km/s, the M-$\sigma$ relation is still less reliable due to the larger uncertainty in the empirical mass-luminosity (M-L) relation at the lower end.  To derive the M-$\sigma$ relation, the dynamical mass of the BHs were estimated using the virial relation, where the virial radius is either measured directly from reverberation mapping (e.g. \citealt{1993PASP..105..247P}) or indirectly from the empirical M-L relation that was derived from the reverberation mapped AGN.  At the low-mass end, few  AGN have been reverberation mapped, so an extrapolation of the M-L relation has to be made in the BH mass estimations, introducing extra uncertainty to the resulting M-$\sigma$ relation.  The M-$\sigma$ relation might be flatter at the lower-mass end based on a few intermediate mass BHs in the sample (e.g. \citealt{2006ApJ...641L..21G}), although the flattening is inconclusive without reverberation mapping results of the lower-mass BHs.}

We validate our BH mass estimation result by comparing to the empirical BH-galaxy mass correlation.  \citet{2015ApJ...813...82R} measure the correlation between galaxy stellar mass and BH mass based on a sample of 341 AGN host galaxies, including a sub-sample of dwarf galaxies \citep{2013ApJ...775..116R}.  They find that $\rm log\left (M_{BH} / M_{\odot}\right) = 7.45 + 1.05 ~\rm log\left(M_*/10^{11} M_{\odot} \right)$, with a \textcolor{black}{scatter} of 0.55 dex.  A stellar mass of \textcolor{black}{$10^8 M_{\odot}$ gives an AGN BH mass of $10^{4 \sim 5} M_{\odot}$} (1$\sigma$ uncertainty assuming log-normal mass distribution), consistent with our BH mass estimation.

\subsection{Keck/LRIS Low Resolution Spectral Analysis}
\label{subsec: lris_low}

\begin{figure}
\centering
    \includegraphics[width=0.46\textwidth]{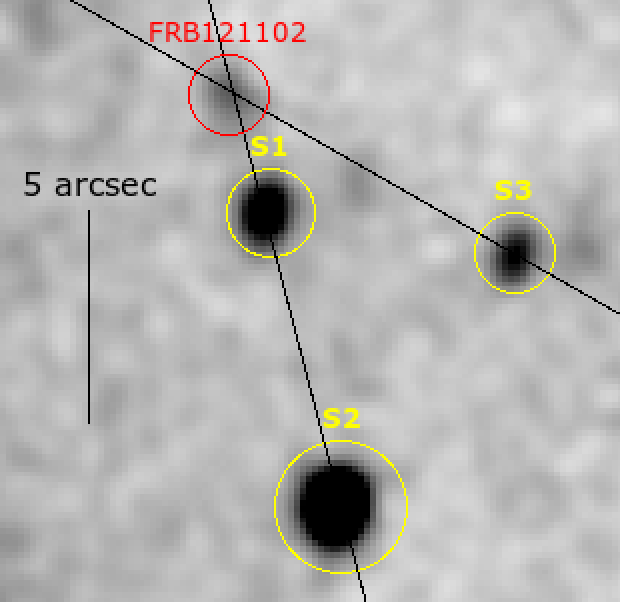}
    \caption{Locations of the host, the slits and the sources in the low resolution Keck/LRIS observation.  The slit orientations are shown as black lines. Spectra were extracted for each of sources S1, S2, and S3, together with the FRB 121102 host. The background image is an LRIS $R$-band exposure, as presented in \citet{2017ApJ...834L...7T}}
    \label{fig:slit}
\end{figure}

Data were obtained with two longslit orientations (shown in Fig. \ref{fig:slit}). The LRIS data were bias subtracted, flattened, cosmic-ray removed, skyline subtracted and flux calibrated using the LPipe \citep{2019PASP..131h4503P}.  The LPipe wavelength calibration failed on the red side, since the arc lamps spectrum missed too many of the expected reference lines.  We calibrated the wavelength manually by fitting the sky spectra to six isolated, bright skylines.  We then selected all nearby sources for which reliable spectra could be extracted-- two sources near the host on slit 1 (S1, S2) and one on slit 2 (S3) (Fig. \ref{fig:slit}).

\textcolor{black}{In this section, we evaluate whether or not the host and its nearby sources belong to the same galaxy group.  The redshift of each spectrum was estimated as follows.  The procedures were also tested on several SDSS spectra and yield results within 1\% of the known redshift values.}
\begin{enumerate}
  \item Remove strong sky line features and data near the \textcolor{black}{boundaries of the wavelength coverage.} 
  \item Interpolate and smooth the spectrum using a \textcolor{black}{1D Gaussian kernel ($\sigma=3$ \r{A}) to reduce random high-frequency noise.  The kernel width was chosen such that it is narrower than the spectral line width at 8000 \r{A} ($1 \sigma \sim 5$ \r{A}) for a typical galactic velocity dispersion of $\sim 200$ km/s (e.g. \citealt{2012PASJ...64..136H}).  Somewhat different kernel width choices (1\r{A}, 3\r{A}, 5 \r{A}) do not change the redshift results significantly.} 
  \item Fit for the continuum and subtract from the spectrum.  \textcolor{black}{The fit was done using the \textit{Astropy} package \textit{specutils}, which removes spikes using a median filter {and fits} the spike-removed spectrum with a list of models using the Levenberg-Marquardt algorithm.}
  \item Extract a 60-\r{A}-long segment of spectrum centered at each typical galaxy absorption (or emission) line and stack them.  \textcolor{black}{We considered the following absorption lines as they are near the visible wavelength at a redshift of $\sim$ 0 to 1: the Balmer series, Ca K\&H (3934.777, 3969.588), the G-band (4305.61), Mg (5176.7), Na (5895.6), and Ca[II] (8500.36, 8544.44, 8664.52).  We ignored emission lines since none of our sources show significant emission features.  The segment width of  60 \r{A} was chosen such that it were able to include at last the $\pm 5 \sigma$ region of a spectral line broadened by a typical galactic vlocity dispersion of 200 km/s (e.g. \citealt{2012PASJ...64..136H}), but not too wide to contain multiple lines.} 
  \item Compare the amplitude of the stacked segment at a grid of \textcolor{black}{trial} redshift values.  In this work, the best estimation occurs at the deepest valley since we only considered absorption lines. 
\end{enumerate} 
  
  \textcolor{black}{For S1, we find $z \approx 0.5796$ and inspect absorptive features at the wavelengths of the Ca H\&K, Mg, H$\gamma$ and possibly the G-band.  For S2, we find $z \approx 0.4471$ and see absorptive features at the wavelengths of the Ca K, the G-band, Na, and potentially H$\alpha$.}  For S3, a few weak lines \textcolor{black}{(the Ca H, Mg and Na)} indicate that this source might be close to S2 in redshift, but we were unable to reach a conclusion due to the low SNR.  We find no significant spectral feature at the expected wavelengths assuming that S3 were at the host redshift.  None of the three sources is likely to be at the same redshift as the host of FRB 121102.

\section{Discussion}\label{sec:discussion}
  \subsection{\textcolor{black}{Does QRS121102 scintillate?}}
  \label{subsec:disc_radio}
  
  \subsubsection{Scattering theory predictions}
  \label{subsubsec:disc_radio_scattering}
  Compact sources scintillate as their wavefronts propagate through \textcolor{black}{an} inhomogeneous ionized medium.  Within the medium, fluctuations in the electron density lead to variations of the refractive index, which change the phase of the wavefront.  The fluctuations can be described by the phase structure function defined as the phase difference of two points separated by a distance $x$: 
  \begin{equation}
  \label{eqn:D}
    D_{\varphi}(x) = \langle[\varphi(x+x_0) - \varphi(x)]\rangle_{x_0} \propto x^{\alpha}.
  \end{equation}
  Here, $\alpha$ is 5/3 for Kolmogorov turbulence \citep{1995ApJ...443..209A}.  We adopt a Kolmogorov turbulence assumption in our calculations thereafter.  When there are relative motions between the source, the medium and the observer, the fluctuations cause temporal variations in the observed flux density. 
  
  For extragalactic sources, \textcolor{black}{scattering is dominated by the Milky Way ISM}, which can often be approximated as a thin scattering screen at a distance $D$ from the observer (e.g., \citealt{1986MNRAS.220...19R}).  One characteristic property of the screen is the  Fresnel scale (e.g. \citealt{1992RSPTA.341..151N}), 
  \begin{equation}
  \label{eqn:r_F}
       r_{\rm F} = \sqrt{\frac{\lambda D}{2 \pi}},
  \end{equation}
  which is the transverse length at which the phase of a wavefront with wavelength $\lambda$ changes by one radian due to the geometric path length difference, assuming that $D \gg \lambda$.  The corresponding angular \textcolor{black}{radius} of the first Fresnel zone is given by  
  \begin{equation}
  \label{eqn:theta_F}
       \theta_{\rm F} = \sqrt{\frac{c}{2 \pi \nu D}}.
  \end{equation}
  Here, $c$ is the speed of light and $\nu$ is the wave frequency.  Another feature of the scattering screen is $r_0$, the transverse scale at which the phase changes by one radian due to the \textcolor{black}{ISM free electron inhomogeneities}.  Based on the relation of these two scales, scattering is divided into the weak regime ($r_0 \gg r_F$) and the strong regime ($r_0 \ll r_F$).  The transitional frequency, $\nu_0$, is defined as the frequency at which \textcolor{black}{$r_0\sim r_F$} for an extragalactic source.  We estimate that  $\nu_0 = 38$ GHz along the line of sight of FRB121102 ($l \approx 175^o$,  $b\approx -0.2^o$) using the NE2001 electron density model (\citealt{2002astro.ph..7156C,2003astro.ph..1598C}).  Our observations were taken at frequencies (12 to 26 GHz) below $\nu_0$, so they all belong to the strong scattering regime.   
  
  In the strong scattering regime, there are two main types of scintillation behaviors: refractive and diffractive.  \textcolor{black}{We summarize the predicted scintillation behaviors below based on \citet{1986MNRAS.220...19R} and \citet{1998MNRAS.294..307W}.}
  
  \textcolor{black}{Diffractive scintillation is produced by interference effects between light rays passing through small-scale ($\ll r_F$) ISM inhomogeneities.  The variations are fast ($t_{\rm d} \sim 2 (\nu/\nu_0) ^{6/5} \sim 5$ hours at $\nu = 18$ GHz) and narrow-band ($\Delta \nu \approx \nu (\nu/\nu_0) ^{17/5} \approx$ 1 GHz at $\nu = 18$ GHz). } 
  
  \textcolor{black}{In contrast, refractive scintillation is caused by large-scale ($\gg r_F$) ISM inhomogeneities}.  The observed flux density variability is slow and broad-band.  For Kolmogorov turbulence, the  \textcolor{black}{angular radius of the} apparent scattering disc at frequency $\nu$ is 
  \begin{equation}
      \label{eqn:theta_r}
      \theta_{\rm r} = \theta_{\rm F0} \left( \frac{\nu_0}{\nu} \right)^{11/5}.
  \end{equation}
  Here, $\theta_{\rm F0}$ is the angular size of the first Fresnel zone at the transitional frequency $\nu_{\rm 0}$.  The observed flux density of a compact source smaller than $\theta_{\rm r}$ varies \textcolor{black}{on} a refractive time-scale (in hours) of    \begin{equation}
      \label{eqn:t_r}
      t_{\rm r} \sim 2 \left(\frac{\nu_0}{\nu} \right) ^{11/5},
  \end{equation}
  assuming a typical relative transverse velocity of 50 km/s \citep{1995A&A...293..479R}.  In this work, modulation index is defined as the weighted root-mean-square (rms) fractional variation: 
  \begin{equation}
  \label{eqn:mp_def}
      m_p = \frac{1}{\langle f \rangle}\sqrt{\frac{\sum_i w_i (f_i - \langle f \rangle)^2}{\sum_i w_i}}. 
  \end{equation}
  Here, $f_i$ is the flux density of the $i$-th epoch, $w_i = 1/{\sigma^2_{f_i}}$ is the weight calculated from the measurement uncertainty $\sigma_{f_i}$, and $\langle f \rangle$ is the weighted mean flux density. The modulation index of a source smaller than $\theta_{\rm r}$ is given by  
  \begin{equation}
      \label{eqn:mp_r}
      m_{\rm p} = \left(\frac{\nu}{\nu_0} \right) ^{17/30}.
  \end{equation}
  \textcolor{black}{When the point source approximation fails ($\theta_s > \theta_r$, where $\theta_s$ is the source angular radius), the modulation index reduces as $m = m_p(\theta_r / \theta_s)^{7/6}$ and the variability timescale increases as $t = t_r(\theta_s / \theta_r)$.}
  
  Table \ref{ta:predictions} lists the predicted $\theta_{\rm r}$, $m_{\rm p}$ and $t_{\rm r}$ using the central frequency of each band, and assuming a point source. We do not predict the substantially larger effects of diffractive scintillation because, as will be seen below, we observe significantly less modulation than expected due to refractive scintillation alone. We calculate $\theta_r$ (eqn. \ref{eqn:theta_F}, \ref{eqn:theta_r}) using a distance $D$ ranging from 100 pc to 10 kpc (Fig. \ref{fig:Tb}), and list the $\theta_{\rm r}$ corresponding to a nominal distance of 1 kpc \textcolor{black}{($\sim$ galactic scale height)} in Table \ref{ta:predictions}.  For scattering \textcolor{black}{media} dominated by a steeper fluctuation spectrum, the expected values of $m_{\rm p}$, $t_{\rm r}$ and $\theta_r$ would be greater \citep{1995ApJ...443..209A}. 

\begin{deluxetable}{c c c c}
\tablecaption{Predicted Galactic Refractive Scintillation Properties of QRS121102 Assuming a Point Source and Kolmogorov Turbulence
\label{ta:predictions}}

 \tablehead{
  \colhead{Band ($\nu_{\rm c}$)} & \colhead{$m_{\rm p}$} & \colhead{$t_r$} & \colhead{$\theta_{\rm r}$ \tablenotemark{a}} \\
  \colhead{} & \colhead{} & \colhead{(Hour)} & \colhead{($\mu as$)}
 }
 \startdata
  K (22 GHz) & $73\%$ & $\sim 7$ & \textcolor{black}{4} \\ 
  U (15 GHz) & $59\%$ & $\sim 16$ & \textcolor{black}{10} \\ 
 \enddata 
 \tablenotetext{a}{Assuming a nominal distance between the scattering screen and the observer to be 1 kpc.}
\end{deluxetable}

  \subsubsection{Comparison between the flux density measurements and predictions} 
  \label{subsubsec: compare_lcv}
  In the following analyses, we analyze two types of flux density measurements in each band-- 
  
  \begin{itemize}
 
    \item The normalized flux density measurements of QRS121102 (``$f3$'' in Table \ref{ta:result_VLA_stat}). 
    
    \item The flux density measurements of QRS121102 divided by those of J0518+3306 and then normalized (``$f3 / f2$'' in Table \ref{ta:result_VLA_stat}).  \textcolor{black}{The scaling is justified for two reasons.  (1) The phase calibrators J0518+3306 (and J0518+3306) are not expected to scintillate, since the constraints on their angular radii and the observed flux densities would require a brightness temperature exceeding the inverse Compton catastrophe threshold -- $\gtrsim 10^{12}$ K (and $\gtrsim10^{13}$ K)-- assuming a scattering screen at 1 kpc.  (2) The variations are unlikely to be intrinsic due to the strong correlations and the short variation timescale, as explained in  Section \ref{subsec:results_radio}.}

  \end{itemize} 
  
   We calculate the modulation indices of the above four sets of flux density measurements (Table \ref{ta:result_VLA_stat}, column 3) and find each of them more than 5$\sigma$ lower than the predictions \textcolor{black}{(error bar calculated from the statistical errors in the flux density measurements)}, though our observation spacings are longer than the predicted refractive time scale (Table \ref{ta:predictions}).  To conclude the comparison, we perform $\chi^2$ tests for two hypotheses -- 
  
  \begin{enumerate}
  
    \item The flux density is constant. 
    
    \item The flux density measurements are drawn from a Gaussian distribution whose modulation is equal to the scattering theory prediction (Table \ref{ta:predictions}, Section \ref{subsubsec:disc_radio_scattering}).  \textcolor{black}{A Gaussian distribution is used here to provide a conservative test, though 
    galactic scattering has been observed to modulate intensity with one-side exponential functions (e.g. \citealt{2002ASPC..278..227C}).} 
    
  \end{enumerate} 
  
  \textcolor{black}{We test the first hypothesis by fitting our measurements in each band with their weighted average and calculate the $\chi^2$:
  \begin{equation}
      \chi^2 = \sum_i\frac{(f_i - \langle f \rangle_w)^{2}}{\sigma_{f_i}^{2}}.
  \end{equation}
  Here, $\langle f \rangle_w$, $f_i$ and $\sigma_{f_i}$ are the weighted average flux density, the i-th epoch flux density and its measurement error, respectively.}  For example, in the K-band, the best-fit result of $f3/f2$ has a $\chi^2_{\rm min}$ of 5.8 and a degree of freedom (dof) of 2, yielding a one-side P-value of $5.6\%$ for obtaining a $\chi^2_{\rm min}$ that is greater or equals to our observation if the flux density were constant.  Other results are listed in Table \ref{ta:result_VLA_stat}.  In both bands, the constant flux density hypothesis is questionable using the $f3/f2$ light curve, and is rejected to a level of at least $10^{-3}$ using the $f3$ light curve. 

  We test the second hypothesis by simulating $10^5$ light curves for both bands, \textcolor{black}{each with the same number of measurements and the same fractional uncertainties as our observations.  For each light curve, the flux densities are drawn from the absolute values of a Gaussian distribution centered at unity and with a standard deviation of $m_p \sqrt{n / (n - 1)}$, where $m_p$ is the expected modulation index and $n$ is the number of measurements.}  We calculate the $\chi^2_{\nu}$ for each light curve and compare the smoothed distribution with our observations in each band.  For example, using $f3/f2$ in the \textcolor{black}{Ku-band}, $\sim 3\%$ of the simulated light curves have $\chi_{\nu}$ values lower than or equals to our observation (2.9), questioning the second hypothesis.  In both bands, the scintillation-variability-hypothesis is  doubtful using the $f3/f2$ data.

\begin{deluxetable*}{c c c c c c}
\tablecaption{VLA Flux Density Modulation Indices Results and Statistical Tests
\label{ta:result_VLA_stat}}

 \tablehead{
  \colhead{Data (band)} & \colhead{$m_{\rm p}'$} & \colhead{$\chi^2_{\rm min}$ (dof)} & \colhead{Constant} & \colhead{Refractive Scintillation}  \\ 
  \colhead{} & \colhead{(Observed)} & \colhead{} & \colhead{$P(\geq \chi^2_{\rm min}; \nu)$} & \colhead{$P(\leq \chi^2_{\nu}; m_{\rm p})$} 
  } 
 
 \startdata
  $f3 / f2$\tablenotemark{a} (K) & $(19.3 \pm 7.1)\%$ & 5.8 (2) & $5.6\%$ &  $\sim 9\%$  \\ 
  $f3 / f2$ (U) & $(13.3 \pm 4.8)\%$ & 10.9 (3) & $1.2\%$ & $\sim 3\%$  \\ 
  \tableline
  $f3$\tablenotemark{b} (K) & $(30.2 \pm 5.8)\%$ & 13.9 (2) & $10^{-3}$ & $\sim 20\%$ \\ 
  $f3$ (U) & $(26.1 \pm 4.4)\%$ & 48.5 (3)  & $10^{-10}$ & $\sim 23\%$ \\ 
 \enddata 
  \tablenotetext{a}{The flux density measurements of QRS121102 divided by those of J0518+3306 and then normalized to an average of unity.}
  \tablenotetext{b}{The normalized flux density measurements of QRS121102.}
\end{deluxetable*} 
  
  \subsubsection{\textcolor{black}{Implications}}  
  \begin{figure}
    \includegraphics[width=0.5\textwidth]{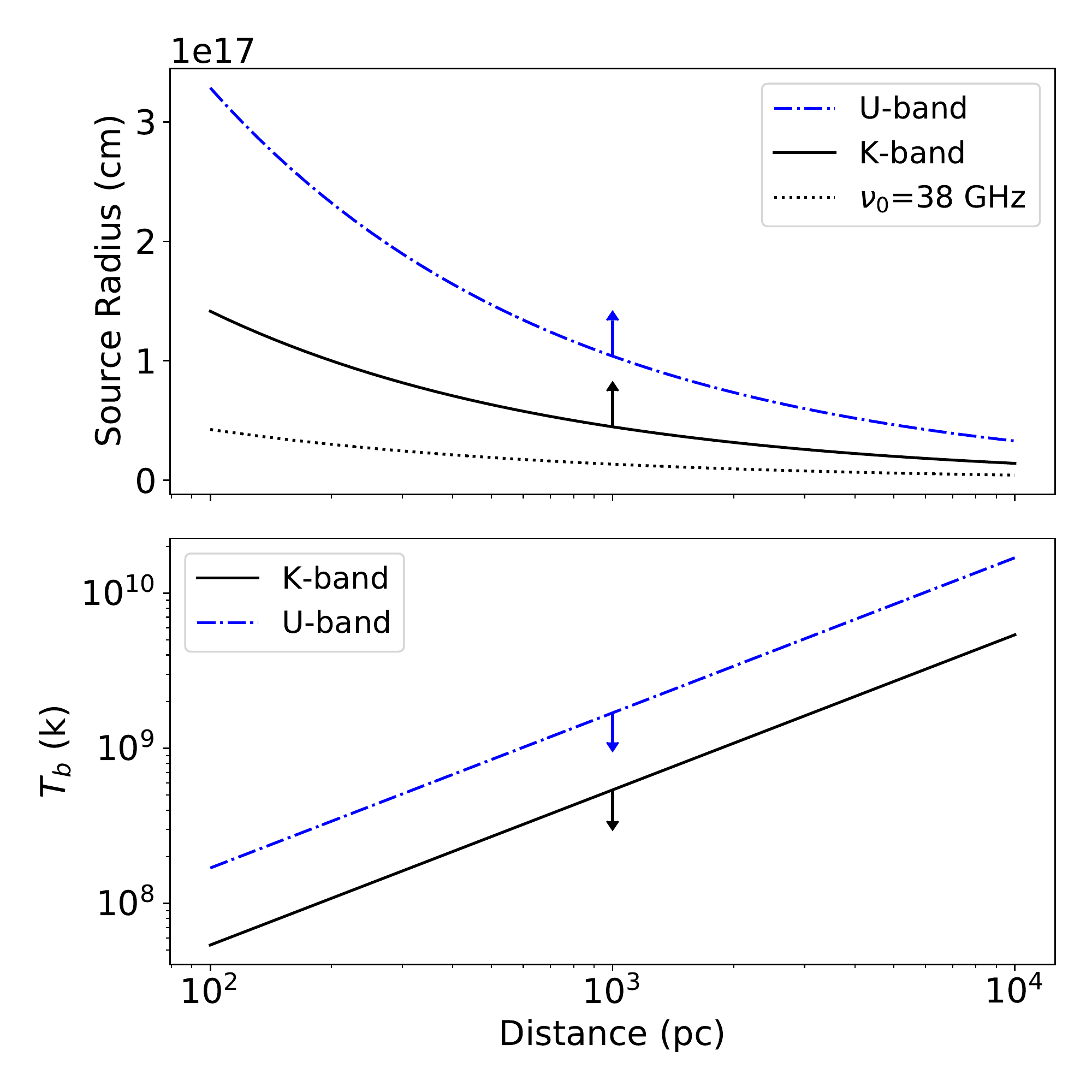}
    \caption{Implication from the lack of refractive scintillation modulation assuming a range of distances between the scattering screen and the observer.  Upper panel: source radius lower limits in the \textcolor{black}{Ku-band} (dashed blue line) and the K-band (solid black line), as well as the size of the scattering disc at the transitional frequency $\nu_0$.  Lower panel: brightness temperature upper limits using the emission region size lower limit inferred from the \textcolor{black}{Ku-band} ($\theta_r \gtrsim 10 ~\mu as$) and the weighted average flux densities measured in the \textcolor{black}{Ku-band} (dashed blue line) and the K-band (solid black line). 
    \label{fig:Tb} }
\end{figure}
  
  \begin{deluxetable*}{c c c c c}
 \tablecaption{Implications of the lack of refractive scintillation. 
 \label{ta:size_Tb}}

  \tablehead{
   \colhead{Observation} & \colhead{$\langle f_{\nu} \rangle$ \tablenotemark{a}} & \colhead{Radio Luminosity \tablenotemark{b}} & \colhead{Radius\tablenotemark{c}} & \colhead{$T_{\rm b}$\tablenotemark{c}} \\
  \colhead{} & \colhead{($\mu$Jy)} & \colhead{(erg s$^{-1}$ )} & \colhead{(cm)} & \colhead{(k)} 
 }
 
 \startdata
  VLA K-band (22 GHz) & $61.3 \pm 5.0 $ & $1.5 \times 10^{39}$ & $\gtrsim 10^{17}$ & $\lesssim 5 \times 10^8$ \\
   VLA \textcolor{black}{Ku-band} (15 GHz) & $89.4 \pm 3.4 $ & $1.5 \times 10^{39}$ & $\gtrsim 4 \times 10^{16}$ & $\lesssim 2 \times 10^9$ \\
 \enddata 
\tablenotetext{a}{Weighted Average Flux Density}
\tablenotetext{b}{Isotropic Luminosity $L \approx \nu_c \langle f_{\nu} \rangle 4\pi D_L^2$, where $\nu_c$ is the band central frequency. }
\tablenotetext{c}{Assuming a nominal distance of 1 kpc between the scattering screen and the observer.}
\end{deluxetable*}

\textcolor{black}{We have} found it questionable that our observation is consistent with refractive scintillation predictions for a point-like source.  We will discuss the implications of the absence of refractive scintillation modulation in our observations.  
  
\textcolor{black}{First, scintillation variations would be smeared out if the angular size of the source is greater than that of the scattering disc.  The lack of modulation implies a lower limit to the physical size of the source in each band (Fig. \ref{fig:Tb}).  Here, the scattering disc angular size lower limit is calculated assuming a range of distances from 100 pc to 10 kpc between the scattering screen and the observer.  The physical size limit of the source  is obtained using the angular size limit and the host redshift ($z=0.19273$)}.  In the \textcolor{black}{Ku-band}, the scattering disc radius is \textcolor{black}{$10~\mu as$} (Table \ref{ta:predictions}) assuming a scattering screen distance of 1 kpc \textcolor{black}{($\sim$ galactic scale height)}, corresponding to a physical radius of $R \gtrsim 10^{17}~\rm cm$ (0.03 pc) at the host redshift.  This together with the VLBI resolution at 5 GHz limit the emission radius to be between $\sim 10^{17}$ cm and $\sim 10^{18}$ cm.   \textcolor{black}{Alternatively, a rough estimation on the source radius could be made using the measured modulation index.  When the angular radius of the source is larger than that of the refractive scattering disc, the variability would reduce and the source radius is given by $\theta_s \approx \theta_r (m_p / m_p')^{6/7}$, where $m_p$ is the expected scintillation modulation index, $m_p'$ is the observed modulation index and $\theta_r$ is the refractive scintillation disc angular radius.  This gives a source radius of 14 $\mu$as ($1.4\times10^{17}$ cm) in the K-band and 36 $\mu$as ($3.7\times10^{17}$ cm) in the Ku-band (using the host redshift and assuming a scattering screen at 1 kpc), within the above constraints of $10^{17\sim18}$ cm.} 

In addition, the constraint on source size rules out the scenario that the flux density modulation is intrinsic.  \textcolor{black}{The flux density varies significantly within a week in the Ku-band (e.g. epochs 3, 4, 5).  If the modulations were intrinsic, the source radius would be $R \lesssim (1/2) \cdot (7~\rm days) \cdot c \sim 10^{14}$ cm, three orders of magnitude below the scintillation size limit.  The source would also have scintillated more if the fast variations were intrinsic.}

  
Moreover, \textcolor{black}{the brightness temperature of the source can be constrained by its size limit and flux density measurement.  Brightness temperature provides clues on the radio emission process.  For example, a brightness temperature above $10^{5}$ K may rule out star-forming galaxies (e.g., \citealt{1992ARA&A..30..575C}), and a brightness temperature above $10^{12}$ K requires coherent processes or relativistic boosting (e.g., \citealt{1979rpa..book.....R}).}  
The brightness temperature is given by: 
 \begin{equation}
  \label{eqn:Tb}
  T_{\rm b} \lesssim \frac{f c^2}{2 \pi \theta_{\rm r, 15GHz}^2 k_{\rm b} \nu^2}.
 \end{equation}
 
Here, $\theta_{\rm r, 15GHz}$ is the radius lower limit implied by the lack of scintillation modulation in the \textcolor{black}{Ku-band}, and $\nu$ is the frequency at which the flux density $f$ is measured.  Assuming a scattering screen distance of 1 kpc (Table \ref{ta:size_Tb}), we find $T_{\rm b} \lesssim 5\times 10^8 ~K$ using our average flux density measured in the K-band ($\nu_c = 22$ GHz, $\langle f_{\nu} \rangle \approx 61.3 \pm 5.0 ~\mu$Jy, \textcolor{black}{weighted by the statistical errors in the flux density measurements}) and $T_{\rm b} \lesssim 2 \times 10^9 ~K$ using the weighted average flux density measured in the \textcolor{black}{Ku-band} ($\nu_c = 15$ GHz, $\langle f_{\nu} \rangle \approx 61.3 \pm 5.0 ~\mu$Jy).  A more complete result assuming a range of scattering screen distances is shown in Fig. \ref{fig:Tb}.



\subsection{Optical} 
\label{subsec:disc_optical}

\subsubsection{\textcolor{black}{Implications of Constraint on Potential supermassive BH Mass}}

\textcolor{black}{We have estimated the mass of the potential supermassive BH to be $10^{4 \sim 5} ~M_{\odot}$ using the velocity dispersion measured from the H$\alpha$ emission line width (section \ref{subsec: lris_high}).  We will test the AGN hypothesis by comparing its mass, radio luminosity and X-ray luminosity with the AGN population.}

\textcolor{black}{The radio luminosity of QRS121102 is several orders of magnitude higher than expected given $M_{BH} \approx 10^{4 \sim 5} M_{\odot}$, based on results from a large sample of AGN with higher BH masses \citep{2001ApJ...551L..17L}. However, it may not be rare in a sample of bright radio emissions detected in dwarf galaxies at intermediate redshifts \citep{2019MNRAS.488..685M}, and is only slightly brighter than those detected in some nearby dwarf galaxies \citep{2020ApJ...888...36R}.  For the general AGN population, the BH mass ($M_{BH}$), radio luminosity ($L_{5 GHz}$), and the ratio of bolometric luminosity to the Eddington limit ($L /L_{Edd}$) has been reported to be correlated.  For example, \citet{2001ApJ...551L..17L} measure a relation based on a sample of 60 AGN with BH masses of $10^{6.5 \sim 10}~M_{\odot}$: $\log_{10}(L_{5 GHz}) = 1.9 \log_{10}(M_{BH} / M_{\odot}) + x \log_{10}(L /L_{Edd}) + 7.9$ (with a scatter of 1.1 dex), where $L_{5 GHz}$ is in a unit of W Hz$^{-1}$ sr$^{-1}$, $x \approx 1$ for a typical $(L /L_{Edd})$ of 0.1 and $x\approx 0.3$ for a low $(L /L_{Edd})$ of $10^{-5}$.  The specific radio luminosity of QRS121102 is  $L_{5 GHz} \approx 10^{21.2}$ W Hz$^{-1}$ sr$^{-1}$, as calculated using the host redshift and the flux density of $\approx 200 ~\mu$Jy at 3GHz and 6 GHz \citep{2017Natur.541...58C}.  This is three orders of magnitude greater than calculated from the relation (scatter included) even if $L /L_{Edd} = 100\%$ and $M_{BH} = 10^5~M_{\odot}$.}   However, we note that the relation reported in \citet{2001ApJ...551L..17L} is derived from a sample of more massive BHs ($10^{6.5 \sim 10}~M_{\odot}$).  

It is intriguing that the radio luminosity of QRS121102 is consistent with a sample of bright radio emissions detected in dwarf galaxies at intermediate redshifts \citep{2019MNRAS.488..685M}.  The specific radio luminosity of QRS121102 at 3 GHz is $L_{3 GHz} \sim 2.3\times10^{22}$ W Hz$^{-1}$, as calculated using the host redshift and the flux density at 3 GHz ($206 \pm 17~\mu Jy$; \citealt{2017Natur.541...58C}).  This lies within the broad luminosity range ($L_{3 GHz} \approx 10^{21.5 \sim 24.2}$ W Hz$^{-1}$) observed in a sample of 35 dwarf galaxies ($10^7 < M_{*} < 10^{9.5}$) at intermediate redshifts (0.13 to 3.4) hosting compact radio sources from the VLA-COSMOS 3 GHz Large Project catalog \citep{2019MNRAS.488..685M}.  These sources are suspected to be AGN mainly because they are significantly more luminous ($\geq 2\sigma$) than expected from star formation processes.  In particular, a few objects in their sample show similar radio luminosities, host stellar masses, BH mass estimations and redshifts with QRS121102, although the constraints on their X-ray luminosities are weaker (Table 1, 2 of \citealt{2019MNRAS.488..685M}).  Similarly, the specific luminosity at 10 GHz ($\sim 1.9\times10^{22}$ W Hz$^{-1}$) is only slightly above the observed range ($10^{18.5 \sim 22}$ W Hz$^{-1}$) from 13 nearby ($z < 0.055$) dwarf galaxies hosting bright radio sources that are too bright to be star formation processes or supernova remnants \citep{2020ApJ...888...36R}.  We suggest that the nature of several of these radio sources, including QRS121102, remains uncertain \citep[e.g.,][]{lca21}.\footnote{Although see \citealt{2021ApJ...910....5M} for an example of an unambiguous AGN in a dwarf galaxy.}

\textcolor{black}{In addition, we estimate the minimum average radio luminosity of QRS121102 during its past life span and find it uncomfortably high given constraints on the X-ray luminosity.  We assume possible source radii of $10^{17}$ cm (lower limit from this work) and $10^{18}$ cm (upper limit from the previous VLBI observation;  \citealt{2017ApJ...834L...8M}) and calculate the minimum energy (equipartition energy) based on section 2.3 of \citealt{2019MNRAS.485L..78V}.  Assuming a power-law electron energy distribution of index -1.5 (to enable direct comparison with \citealt{2019MNRAS.485L..78V} and to account for the flat spectrum), the minimum total energies required to power a synchrontron source with the observed radio luminosity at these two size limits are $E_{\rm q} \approx 10^{48.9}$ ergs ($B_{\rm eq} \approx 27$ mG) and $\approx 10^{50.2}$ ergs ($B_{\rm eq} \approx 190$ mG), respectively (e.g. Chapter 5 of \citealt{2016era..book.....C}).  Adopting a conservative expansion speed  of $\sim 0.01$c \citep{2019MNRAS.485L..78V}, the average radio luminosity during its past lifespan would be $2 \times 10^{40}$ erg s$^{-1}$ ($5 \times 10^{40}$ erg s$^{-1}$), about $0.2\%$ ($0.5\%$) of the Eddington limit for a $10^5 M_{\odot}$ BH. This is uncomfortably high, accounting for typical amounts ($\sim10\%$) of energy deposited into relativistic electrons, given the upper limit on X-ray emission of 4\% of the Eddington limit for a $10^5 M_{\odot}$ BH \citep{2017Natur.541...58C}. }

\textcolor{black}{The radio and X-ray observations of QRS121102 can also be compared with the radio / X-ray luminosity correlation in accreting BH systems (e.g. \citealt{1998A&A...337..460H}).  In particular, the AGN BH mass has been found to be correlated with its radio and X-ray luminosity (e.g. \citealt{2004A&A...414..895F, 2003MNRAS.345.1057M}).  A recent report based on a sample of 30 AGN with independent dynamical mass measurements shows that $\rm log\left(M/10^8~M_{\odot}\right) = 0.55 + 1.09~ \rm log\left(L_R/10^{38} \rm erg~s^{-1}\right) -0.59~ \rm log\left(L_X/10^{40} \rm erg~s^{-1}\right)$ \citep{2019ApJ...871...80G}, with a 1$\sigma$ \textcolor{black}{scatter} of 1 dex assuming a log-normal mass distribution.  Here, $L_R$ and $L_x$ are the luminosity at 5 GHz and 2 to 10 keV, respectively, observed within close epochs ($\Delta t \lesssim 2+M/10^6 M_{\odot}$ days).  We adopt a 5 GHz flux density of $f_R \approx 5~\rm GHz \cdot 200 \mu \rm Jy \sim 10^{-17} \rm erg~s^{-1}~cm^{-2}$ measured by VLA and an X-ray flux upper limit of $f_x \lesssim 5\times10^{-15} \rm erg~s^{-1}~cm^{-2}$ inferred from the non-detection in the XMM-Newton and Chandra images \citep{2017Natur.541...58C}.  We convert the flux density to isotropic luminosity using the host redshift and have $L_R \approx 10^{39} \rm erg~s^{-1}$ and $L_x \lesssim 5 \times 10^{37} \rm erg~s^{-1}$, giving a BH mass of $\sim 10^{11} ~M_{\odot}$, $6 \sim 7$ orders of magnitude greater than our measurement.  Therefore the persistent radio source does not follow the AGN BH mass-luminosity relation measured in \citet{2019ApJ...871...80G}. }

\subsubsection{AGN in an Isolated Dwarf Galaxy}

\textcolor{black}{A fraction (8\% to 32\%) of supermassive BH have been estimated to reside within high mass ultra-compact dwarf galaxies, suggesting that some of those galaxies could be the stripped cores of larger galaxies through tidal interactions with their companions \citep{2019ApJ...871..159V}.  We find that this scenario is not supported for QRS121102 for two reasons.  First, the low BH mass estimated from the gas velocity dispersion is consistent with a typical dwarf galaxy instead of a more massive galaxy (section \ref{subsec: lris_high}).}  \textcolor{black}{Second, member(s) from the same galaxy group are expected to be associated with QRS121102 if the host had been dynamically stripped by nearby companions.  From the low resolution LRIS spectra, we found that the three nearby bright sources are likely to have different redshift values from the host (section \ref{subsec: lris_low}).}  

We extend this argument by searching the PanSTARRS catalog for potential companions that are likely to belong to the same galaxy group. \textcolor{black}{We search the PanSTARRS catalog for sources within 5 arcmin ($\sim$1 Mpc, the virial radius of a galaxy group with a typical mass of $\sim 10^{13}~M_{\odot}$ and velocity of 200 km/s) and have consistent photometric redshift measurements with the host.  One object (PSO J082.9850+33.0967) was found at 3' from QRS121102 but only detected in a stacked image and have no valid magnitude measurement available in the catalog.  Another object (PSO J082.9961+33.0895) was ignored due to the large uncertainty in its photomatric redshift ($0.20 \pm 0.18$).  We find no promising group member candidate from the PanSTARRs catalog.} {Moreover, we compare the PanSTARRS galaxy number density within this area with the galaxy number density calculated from the deep VRI images in the R-band produced by the Keck Telescope \citep{1995ApJ...449L.105S}.  In that work, they estimate a galaxy number density of $\approx 7 \times 10^5$ deg$^{-2}$ with a magnitude range of $20.5 \lesssim m_R \lesssim 27.2$.  In the PanSTARRS DS1 catalog, 2705 objects are found within a radius of 5 arcmin around FRB121102, and 30 of them are classified as galaxies above a confidence level of 90\% \citep{2018PASP..130l8001T}.  The limiting magnitude of PanSTARRS is $m_r \approx 23.2$, six times shallower than that of the deep VRI images, predicting $\approx 60$ galaxies within the searched area at a limiting magnitude of 23.2.  The galaxy number density near QSR121102 is not overdense compare to an average sky region.  We find no evidence that the host belongs to a galaxy group.}

\subsection{\textcolor{black}{What Else Could the Source Be}} 

\textcolor{black}{We have found that QRS121102 is unlikely to be an AGN based on the low inferred BH mass ($\lesssim 10^{4\sim 5} M_{\odot}$), high radio-to-X-ray luminosity ratio and the absence of companions from the same galaxy group.  In this section, we discuss other possible sources for the compact persistent radio emission.}

The size of the persistent radio emission could be explained by an isolated young neutron star with luminous synchrotron emission produced in a pulsar wind nebula (PWN), or plerion.  The pulsar wind forms a terminal shock at a radius where the wind momentum flux and the confining pressure reach a balance, and forms a PWN further out.  The shock radius is given by: $r_w = \sqrt{\Dot{E} / (4 \pi \eta c p)}$ \citep{2005AdSpR..35.1092S}, where $\Dot{E}$ is the pulsar energy injection rate into the wind, $\eta$ is the fraction of area covered by the wind, $c$ is the speed of light, $p$ is the confining pressure outside the shock and is proportional to the electron number density $n_e$ for medium with the same components.  The relation gives a radius of $\sim 0.1$ pc for a canonical isolated radio pulsar, {and is confirmed to be $\approx 3\times 10^{17}$ cm from the X-ray images of the Crab nebula \citep{2000ApJ...536L..81W}.  For the Crab pulsar, the spin down energy rate is $\Dot{E} \sim 4.5\times10^{38}$ ergs s$^{-1}$ \citep{1968Sci...162.1481S}, and the confining pressure outside the shock is proportional to the density of the medium, which can be approximated as a typical ISM ($n_{e} \sim 10^{-1}$ to $10^{-2}$ cm$^{-3}$; e.g. \citealt{2011piim.book.....D}).  In comparison, the energy ejection rate of FRB 121102 by the flares and the wind into the surrounding medium is estimated to be $\Dot{E} \sim 10^{39}$ to $10^{40}$ ergs s$^{-1}$ (Fig. 5 of \citealt{2020MNRAS.494.4627M}), and the medium ahead of the termination shock is likely denser than a typical ISM, as indicated from the high RM of FRB 121102 \citep{2018Natur.553..182M}}.

One example of the PWN emission model \textcolor{black}{that produce the observed radio luminosities and the size of QRS121102} is presented in \citet{2018ApJ...868L...4M}, where the authors explain the persistent emission using a concordance FRB model.  On a large scale ($\sim 10^{15}$ cm, Eqn. 4 of \citealt{2019MNRAS.485.4091M}), the train of ion-electron shells merge into a steady wind and feed into a nebula via a terminal shock, which heats up electrons in the nebula and produces the persistent synchrotron radio source.  Based on the lack of self-absorption feature down to 6.0 GHz in the spectrum of the persistent source \citep{2017Natur.541...58C}, \citet{2018ApJ...868L...4M} estimate that the emission region's radius $R \gtrsim 0.46 \times 10^{17}~\rm cm$ \textcolor{black}{adopting the observed luminosity at 6 GHz}.  Moreover, \citet{2020arXiv201014334R} report the lack of self-absorption down to 400 MHz in their GMRT observations.  Using their flux density measurement at 400 MHz and the scaling relation $R \propto L_{\nu, \rm obs}^{4/11} \nu_{\rm obs}^{-10/11}$ (Eqn. 21 in \citealt{2018ApJ...868L...4M}), we find that $R \gtrsim 5.4 \times 10^{17}~\rm cm$.  This emission size is consistent with the constraint implied by the lack of refractive scintillation in the \textcolor{black}{Ku-band}. 

Among the more unique features of QRS121102 is its unusually flat radio spectrum at GHz frequencies \citep{2017Natur.541...58C}. As we have noted, several examples of compact radio sources of similar luminosities exist in dwarf galaxies. Indeed, two recent discoveries of \textit{transient} radio sources not associated with BHs also reached similar radio luminosities. The first, FIRST J141918.9+394036, had a peak radio luminosity of $2 \times 10^{29} \rm~erg~d^{-1}~Hz^{-1}$ at 1.4 GHz, and is hosted by a star-forming dwarf galaxy \citep{2018ApJ...866L..22L}. FIRST J141918.9+394036 is most likely the afterglow of an off-axis long-duration GRB \citep{mml+21}. The second,  VT J121001+495647, had a peak radio luminosity of $1.5 \times 10^{29} \rm ~erg~s^{-1}~Hz^{-1}$ at 5 GHz, was associated with a star-forming region, and was ascribed to interaction with a dense circum-stellar medium ejected through binary interaction \citep{2021Sci...373.1125D}. In both cases, however, classical synchrotron spectral shapes were observed together with secular time-evolution, unlike in the case of QRS121102. We urge continued wideband monitoring of QRS121102, together with more detailed evaluation of potential empirical analogs.

\textcolor{black}{Finally, we rule out a few other possible origins of QRS121102 based on our observations.  First, the source cannot be a supernova remnant (SNR) due to its high luminosity.  We have calculated the specific luminosity of QSR121102 at 3 GHz and 10 GHz as $L_{\nu} \sim 10^{22}$ W Hz$^{-1}$.  \citet{2019A&A...623A.173V} and \citet{2009AJ....138.1529U} recently report the radio luminosities (5 GHz and 8.4 GHz) of 102 SNRs in the merging galaxies Arp 220 and Arp 229.  The brightest SNR in the sample is $L_{\nu} < 10^{21}$ W Hz$^{-1}$, over one order of magnitude lower than that of QSR121102.  Moreover, the luminosity is inconsistent with the star formation rate (SFR) of the host galaxy if the source were SNR(s).  The brightest SNR and the SFR of a galaxy have been reported to be related \citep{2009ApJ...703..370C} as: $L_{1.4}^{max} = 95^{+31}_{-23} \rm~SFR^{0.98\pm0.12}$, where $L_{1.4}^{\rm max}$ is in a unit of $10^{24}$ erg s$^{-1}$ Hz$^{-1}$ and SFR is in $M_{\odot}$ year$^{-1}$.  Adopting the SFR upper limit of 0.4 $M_{\odot}$ year$^{-1}$ based on the host H$\alpha$ emission flux \citep{2017ApJ...834L...7T}, the brightest SNR in the host would be $L_{1.4}^{\rm max} \approx 10^{18.6}$ W Hz$^{-1}$, over 3 orders of magnitude lower than that of QSR121102.  The persistent source is too bright to be an SNR.} Second, the source is too bright for a long-duration GRB (LGRB) radio afterglow.  Adopting a typical LGRB peak radio luminosity of $L_{\nu, \rm 8.5 GHz} \sim 2\times 10^{31}$ erg$^{-1}$ s$^{-1}$ Hz$^{-1}$ and a decay rate of $\propto t^{-2}$ from day 10 after the GRB (e.g. \citealt{2003Natur.426..154B}), the radio luminosity would have reached the level of QRS121102 ($L_{\nu, \rm 10 GHz} \approx 2 \times 10^{29}$ erg$^{-1}$ s$^{-1}$ Hz$^{-1}$) within 3 weeks, while the radio luminosity of QRS121102 have been nearly constant below 10 GHz for years (e.g. \citealt{2020arXiv201014334R, 2017ApJ...834L...7T}).

\section{Conclusion}
\label{sec:conclusion}

In this work, we investigated the origin of the persistent radio source, QRS121102, associated with FRB 121102. We present new VLA monitoring data (12 to 26 GHz) and new spectra from Keck/LRIS.  The main results are summarized as follows:

\begin{enumerate}
\item We constrained the emission radius to be $10^{17 \sim 18}$ cm based on the low level of scintillation variability in our VLA observations and the previous VLBI observation.  A few compact radio sources (e.g. AGN, PWNs, very young SNRs and GRB afterglows) could fall into these size limits. Most interpretations {with the exception of an AGN} would have been in tension with a converse finding of significant scintillation in QRS121102. 

\item To further investigate the hypothesis that the source is an AGN, we roughly constrained the mass of the potential BH to be $\lesssim 10^{4\sim 5} M_{\odot}$ using the H$\alpha$ velocity dispersion.  The radio luminosity ($L_{\nu} \sim 2 \times 10^{22}$ W Hz$^{-1}$ from 400 MHz to 10 GHz) is possibly too high at this BH mass compared to the general AGN population, although similarly bright radio emissions have been reported in several dwarf galaxies.  The source is also unlikely to be an AGN because it is too faint in the X-ray for its low BH mass and bright radio emission. 

\item A significant fraction of dwarf galaxies hosting supermassive BHs may be the stripped cores of massive galaxies during tidal interactions with their nearby companion(s).  From our LRIS spectra and the PanSTARRS catalog, we found no promising companion galaxy near the host to support an environment for a tidal stripping event history.  

\item We briefly discussed possible origins other than AGN.  QRS121102 is too luminous in the radio band to be an SNR and too old to be a typical GRB afterglow.  The isolated young neutron star models for FRBs might be able to account for both the size and the luminosity of the persistent source as synchrotron emission produced in the PWN (e.g. \citealt{2019MNRAS.485.4091M}).     
\end{enumerate}

In conclusion, the persistent radio emission associated with FRB 121102 is likely not an AGN, and its nature remains interesting for FRB emission models involving extreme neutron stars. 

\acknowledgments
The authors thank Dr. Casey J. Law for VLA imaging and visualization tips and the catalog querying code \textit{psquery}.  We also thank Dr. Casey J. Law and Dr. Liam D. Connor for helpful discussions on PRS and the unknown radio emissions in dwarf galaxies.  We thank staff members at the CASA help desk, Dillon Dong and Nitika Yadlapalli for CASA tips.  This material is based upon work supported by the National Science Foundation under grant No. AST-1836018.

\bibliography{121102_persistent_emission.bib}
\bibliographystyle{aasjournal}

\end{document}